\documentclass[a4paper,12pt,notitlepage]{article}
\usepackage{amsmath}
\usepackage{latexsym}
\usepackage{amssymb}
\usepackage{amsfonts}
\usepackage{verbatim}

\newcommand{\R}{ \mathbb{R} }
\newcommand{\C}{ \mathbb{C} }

\newcommand{\tr}{\mathop{\mathrm{tr}}}
\newcommand{\diag}{\mathop{\mathrm{diag}}}
\newcommand{\re}{\mathop{\mathrm{Re}}}
\newcommand{\im}{\mathop{\mathrm{Im}}}

\newcommand{\sign}{\mathop{\mathrm{sgn}}}
\newcommand{\dd}{  {d^{\, 2}\!}  }

\renewcommand{\vec}[1]{\boldsymbol{#1}}

\newtheorem{Thm}{Theorem}%[section]
\newtheorem{Lemma}[Thm]{Lemma}
\newtheorem{Prop}[Thm]{Proposition}

\numberwithin{equation}{section}

\author{Yan V. Fyodorov\footnote{\noindent School of Mathematical Sciences, University of Nottingham, Nottingham, NG7 2RD,
U.K.; E-mail: yan.fyodorov@nottingham.ac.uk. The research in
Nottingham is supported by EPSRC grant  EP/C515056/1: "Random
Matrices and Polynomials: a tool to understand complexity"}
\hspace*{1ex} and \hspace*{0.5ex}Boris A.
Khoruzhenko\footnote{School of Mathematical Sciences, Queen Mary,
University of London, London E1 4NS, U.K.; E-mail:
b.khoruzhenko@qmul.ac.uk. A significant part of this work was
carried out during the Newton Institute programme on Random Matrix
Approaches in Number Theory (26 January-16 July 2004)}}

\title{On absolute moments of characteristic polynomials of a certain class of complex random matrices}
\date{31 January 2006}

\begin{document}

%\begin{titlepage}

\maketitle

\begin{abstract}
Integer moments of the spectral determinant $|\det(zI-W)|^2$ of
complex random matrices $W$ are obtained in terms of the
characteristic polynomial of the Hermitian matrix $WW^*$ for the
class of matrices $W=AU$ where $A$ is a given matrix and $U$ is
random unitary. This work is motivated by studies of complex
eigenvalues of random matrices and potential applications of the
obtained results in this context are discussed.
\end{abstract}

%\end{titlepage}

\section{Introduction} \label{sec1}

Characteristic polynomials of random matrices have recently
attracted considerable interest in the mathematical physics
literature. Initially, the interest was stimulated by applications
in number theory \cite{KS1,KS2}, quantum chaos \cite{AS,Fyo,FStra}
and quantum chromodynamics (QCD) \cite{SV,V1,HJV,FA}, but with the
emerging connections to integrable systems \cite{O1,V2},
combinatorics \cite{DG,Strahov}, representation theory
\cite{BOS,BS,BG,CFZ} and analysis \cite{BDS}, it has become
apparent that characteristic polynomials of random matrices are
also of independent interest.

In this paper we are concerned with integer moments of the squared
modulus of characteristic polynomial of complex random matrices in
a rather general class of matrices $W=AU$, where $A$ a given
positive semidefinite Hermitian matrix, $A\ge 0$, and $U$ is a
random unitary matrix distributed uniformly over the unitary
group.

In the particular case when $A$ is identity matrix, the matrix $W$
is random unitary, and its eigenvalues lie on the unit circle.
Various moments of the characteristic polynomial for this class of
matrices were obtained recently, see \cite{KS1,KS2,CFKRS,CFS,CFZ}.
In the general case, the eigenvalues of $W=AU$ will be distributed
in a region in the complex plane. Eigenvalue statistics of such
complex eigenvalues, and in particular the mean eigenvalue
density, are of interest for physics of open chaotic systems, see,
e.g. \cite{FSom1,FSom2}, and in QCD, see, e.g. \cite{V2} and
references therein, and are difficult to study analytically. In
this context moments of the squared modulus of the characteristic
polynomial frequently provide a very useful tool.  Indeed, in a
variety of random matrix ensembles the mean eigenvalue density,
\begin{equation}\label{int:eq00}
\rho (x,y) = \langle \tr \delta (zI-W) \rangle_W, \hspace{3ex}
z=x+iy,
\end{equation}
can be expressed in terms of the mean-square-modulus of the
characteristic polynomial in a closely related random matrix
ensemble. In (\ref{int:eq00}) the angle brackets stand for
averaging over the matrix distribution, and $I$ is identity
matrix.

An obvious example is served by the Ginibre ensemble of complex
matrices \cite{G}. In this ensemble the matrix distribution has
density $Const.\times \exp (-\tr WW^*)$ where $W^*$ is complex
conjugate transpose of $W$. The mean density $\rho_n (x,y)$ of
eigenvalues of Ginibre matrices of size $n\times n$ is given by
\begin{equation}\label{int:eq0}
\rho_n(x,y)= \frac{1}{\pi}e^{-|z|^2} \sum_{k=0}^{n-1}
\frac{|z|^{2k}}{k!}.
\end{equation}
One can arrive at (\ref{int:eq0}) in various ways. Ginibre
computed the joint probability density function of all eigenvalues
and then applied the method of orthogonal polynomials. Another way
is to use the method of dimensional reduction, see e.g.
\cite{Trotter,E,EKS}, which yields the following relation
\begin{equation}\label{int:eq1}
\rho_n(x,y)=\frac{e^{-|z|^2}}{\pi (n-1)!}  \langle\left|\det
(zI_{n-1} - W_{n-1}) \right|^2\rangle_{W_{n-1}}.
\end{equation}
Here the angle brackets stand for averaging over the Ginibre
ensemble of complex matrices of size $(n-1)\times (n-1)$. The
mean-square on the r.h.s. in (\ref{int:eq1}) can be easily
computed yielding again (\ref{int:eq0}).

A less obvious example, which in fact provided the initial impetus
for the present study, is the so-called ensemble of `random
contractions' \cite{FSom2}. In its simplest variant of rank-one
deviations from the unitarity, these are random $n\times n$
matrices satisfying the constraint
\begin{equation}\label{con}
W_nW_n^* = \begin{pmatrix}
  1-\gamma & 0 \\
  0 & I_{n-1}
\end{pmatrix}, \hspace{3ex} 0< \gamma <1.
\end{equation}
In the `polar' coordinates, $W_n=G_nU_n$ where $U_n$ is a CUE$_n$
matrix, i.e. a matrix drawn at random from the unitary group $
U(n)$, and $G_n = \diag\ (\sqrt{1-\gamma}, 1, \ldots, 1)$. The
mean density of eigenvalues\footnote{Note that constraint
(\ref{con}) implies that the eigenvalues of $W$ lie in the annulus
$1-\gamma \le |z|^2 \le 1$.} of $W_n$ can be expressed as the mean
square modulus of the characteristic polynomial of $(n-1)\times
(n-1)$ matrices $ \tilde G_{n-1}U_{n-1}$,
\begin{equation}\label{int:eq2}
\rho_n(x,y) = \frac{n-1}{\pi \gamma |z|^2} \left(\frac{\tilde
\gamma}{\gamma} \right)^{n-2} \langle |\det (zI_{n-1} - \tilde
G_{n-1}U_{n-1})|^2 \rangle_{U_{n-1} }, \hspace{3ex} 1-\gamma <
|z|^2 < 1,
%    0 & \text{if $|z|^2>1$ or $|z|^2< 1-\gamma $},
\end{equation}
where now the angle brackets stand for averaging over the unitary
group $U(n-1)$ with respect to the normalized Haar measure, and
\[
\tilde G_{n-1}= \diag (\sqrt{1-\tilde \gamma}, 1, \ldots, 1),
\hspace{3ex} \tilde \gamma = \frac{|z|^2+\gamma-1}{|z|^2}.
\]

Another example is provided by finite-rank deviations from
Hermiticity \cite{FKh}. We only consider the simplest but still
non-trivial case of rank-one deviations. Let
\begin{equation}\label{res}
 W_n=H_n+i\Gamma_n
\end{equation}
where $H_n$ is a GUE$_n$ matrix, i.e. random Hermitian matrix of
size $n\times n$ with probability distribution
\[
dP_{\beta,n}(H)=Const. \times e^{-\frac{\beta}{2} \tr H^2}\
\prod_{j=1}^n dH_{jj} \prod_{1\le j<k\le n}
\frac{dH_{jj}d\overline{H_{jj}} }{2}, \hspace{3ex} \beta
>0,
\]
and $\Gamma_n= \diag(\gamma, 0, \ldots, 0)$, $\gamma >0$. It is
apparent that all eigenvalues of $W$ lie in the strip $0\le y \le
\gamma$. For the mean eigenvalue density $\rho_n(x,y)$ of $W_n$ we
have
\begin{equation}\label{int:eq2h}
\rho_n(x,y) = r_{\beta,n}(x,y) \  \langle |\det (zI_{n-1} -
(H_{n-1}+i\tilde \Gamma_{n-1})|^2 \rangle_{H_{n-1}}, \hspace{3ex}
0<y<\gamma,
\end{equation}
where
\[
r_{\beta,n}(x,y)= \frac{\beta^n(\gamma-y)^{n-2}e^{-\frac{\beta
x^2}{2}-\beta (\gamma-y)y}}{4\sqrt{2\pi\beta}\
\gamma^{n-1}(n-2)!}, \hspace{3ex} \tilde \Gamma_{n-1}=
\diag(\gamma-y,0,\ldots, 0),
\]
and the angle brackets stand for averaging with respect to the
distribution $dP_{\beta,n-1}(H)$. We derive (\ref{int:eq2}) and
(\ref{int:eq2h}) in Section \ref{sec6}.

The above formulas relating the mean eigenvalue density and the
mean-square-modulus of characteristic polynomial are specific to
the considered matrix distributions. In general situation, the
mean density of eigenvalues can be determined from fractional
moments of the squared modulus $|\det (zI-W)|^2= \det
(zI-W)(zI-W)^*$ (e.g., by the way of the logarithmic potential of
the eigenvalue distribution), or from averages of ratios of $\det
[(zI-W)(zI-W)^* + \varepsilon^2 I]$, see e.g. \cite{FSom2}.
Getting explicit formulas for that kind of objects outside the
classes of Hermitian and unitary matrices is, however, a
considerable challenge. Although it is known that the fractional
moments of $|\det (zI-W)|^2$ can be written in terms of a
hypergeometric function of matrix argument $WW^*$ \cite{O1}, the
corresponding series is hard to deal with in the limit of infinite
matrix dimension.

Our main result, Theorem \ref{thm1}, expresses
\[
\langle \left[\det (zI-AU) (zI-AU)^{*}\right]^{\pm m} \rangle_U,
\hspace{3ex} m=1,2, \ldots ,
\]
where the integration is over unitary matrices $U$ with respect to
the Haar measure, as an $m$-fold integral of powers of the
characteristic polynomial of the (Hermitian!) matrix $AA^*$. This
integral can be written an $m\times m$ determinant with entries
given by a certain integral transform of the characteristic
polynomial of $AA^*$, see (\ref{p1})-(\ref{n1}). In particular,
this result implies that for the ensembles of random complex
matrices $W$ with unitary invariant matrix distribution (for
example, for the Feinberg-Zee ensemble \cite{FZ} whose probability
density of matrix entries of $W$ depends only on $WW^*$ ) our
formulas effectively reduce the original non-Hermitian problem to
a Hermitian one, albeit on the level of characteristic
polynomials. This, as explained in more detail at the end of the
next section, has a clear computational advantage, as one can then
use various formulas for averages of products and ratios of
characteristic polynomials of Hermitian matrices which have been
obtained in recent years, see \cite{BH,FS1,BDS}. In contrast, with
the exception of essentially Gaussian weights \cite{AV,AP}, no
such formulas are known for complex matrices.

We also express
\[
\left\langle\frac{1}{\det [(zI-AU)(zI-AU)^* +\varepsilon^2 I ]}
\right\rangle_U
\]
as a two-fold integral of the inverse spectral determinant of
$AA^*$, see Theorem \ref{thm2a}. Again, the non-Hermitian problem
is reduced to a Hermitian one. This regularized inverse spectral
determinant can be useful as an indicator of the domain of the
distribution of complex eigenvalues. For, any point $z$ where the
above average blows up in the limit $\varepsilon \to 0$ must
belong to the domain of eigenvalue distribution.

\bigskip

{\bf Acknowledgments.} We would like to thank A. Gamburd for useful
discussions and, in particular, for bringing reference
\cite{Trotter} to our attention and for pointing us towards the link
between our Lemma \ref{Lemma1} and the Selberg Integral. We are also
grateful to Ph. Biane for bringing reference \cite{HL} to our
attention.

\section{Statement of main results and discussion} \label{sec2}
Let $n$ and $m$ be positive integers. Define
\begin{equation}\label{mu}
d\mu_n(t_1, \ldots, t_m)=  \frac{1}{c_n} \ \Delta^2 (t_1, \dots,
t_m)\ \prod_{j=1}^m (1+t_j)^{-n-2m}\ dt_1 \ldots dt_m,
\hspace{3ex} t_j \ge 0,
\end{equation}
and, for $n \ge 2m$,
\begin{equation}\label{nu}
d\nu_n(t_1, \ldots, t_m)=  \frac{1}{k_n}\  \Delta^2 (t_1, \dots,
t_m)\ \prod_{j=1}^m (1-t_j)^{n-2m}\ dt_1 \ldots dt_m, \hspace{3ex}
0\le t_j \le 1,
\end{equation}
where
\begin{equation}\label{vander}
\Delta (t_1, \dots, t_m) = \det \left(t_i^{m-j}\right)_{i,j=1}^m =
\prod_{1\le i<j\le m} (t_i-t_j)
\end{equation}
is the Vardermonde determinant, and
\begin{equation}\label{cnkn}
c_n=\prod_{j=0}^{m-1} \frac{j!(j+1)!(n+j)!}{(n+m+j)!}
\hspace{2ex}\hbox{and} \hspace{2ex} k_n=\prod_{j=0}^{m-1}
\frac{j!(j+1)!(n-m-j-1)!}{(n-j-1)!}
\end{equation}
are the normalization constants. The Selberg Integral, see e.g.
\cite{Mehta}, asserts that $d\mu_n$ and $d\nu_n$ are unit mass
measures,
\[
\int\limits_0^{\infty} \hspace{-0.5ex}\ldots \hspace{-0.5ex}
\int\limits_0^{\infty}\ d\mu_n(t_1, \ldots, t_m) =\int\limits_0^1
\hspace{-0.5ex}\ldots \hspace{-0.5ex} \int\limits_0^1\ d\nu_n(t_1,
\ldots, t_m) = 1.
\]

The measures $d\mu_n$ and $d\nu_n$ define probability
distributions which have the following random matrix
interpretation. Consider two families of matrix distributions on
the space of $m\times m$ complex matrices
$Z=\left(x_{jk}+iy_{jk}\right)_{j,k=1}^m$:
\begin{equation}\label{muhat}
 d\hat
\mu_{n} (Z)= \frac{1}{\hat c_n}\ \frac{1}{{\det}^{n+2m}
(I_m+ZZ^*)}\ \prod\limits_{j,k=1}^{m} dx_{jk}\ d y_{jk},
\hspace{3ex} n \ge 0,
\end{equation}
and
\begin{equation}\label{nuhat}
d\hat\nu_{n} (Z)=\frac{1}{\hat k_n}\ {\det}^{n-2m} (I_m-ZZ^*)\
 \prod\limits_{j,k=1}^{m} dx_{jk}\ d y_{jk},
\hspace{3ex} n \ge 2m.
\end{equation}
The measures $d\hat\nu_{n} (Z)$ are defined on the matrix ball
$ZZ^*<I_m$ and the constants $\hat c_n$ and $\hat k_n$ are
determined by the normalization condition
\[
\int_{ZZ^* \ge 0} d\hat\mu_n (Z)= \int_{0\le ZZ^* \le I_m }
d\hat\nu_n (Z)=1.
\]
A standard calculation, see e.g. \cite{Hua}, shows that if $s
(ZZ^*)$ is a symmetric function of the eigenvalues $t_1, \ldots ,
t_m$ of $ZZ^*$, i.e. $s (ZZ^*)= s(t_1, \ldots, t_m)$, then
\begin{eqnarray}\label{tt}
\int_{ZZ^* \ge 0} \  s(ZZ^*)\ d\hat\mu_n (Z)&=&
\int\limits_{0}^{+\infty}\hspace{-0.5ex} \cdots
\int\limits_{0}^{+\infty}\  s(t_1, \ldots, t_m)\ d\mu_n(t_1,
\ldots t_m), \\ \label{ss} \int_{0\le ZZ^* \le I_m }\  s(ZZ^*)\
d\hat\nu_n (Z)&=& \int\limits_{0}^{1}\hspace{-0.5ex} \cdots
\int\limits_{0}^{1} \ s(t_1, \ldots, t_m)\ d\nu_n(t_1, \ldots
t_m).
\end{eqnarray}

Theorem \ref{thm1} below, which we state in a slightly more
generality than required for spectral determinants, tells how to
integrate moments of determinants over the unitary group equipped
with the Haar measure $dU$ fixed by the normalization $
\int_{U(n)} dU=1$.

\bigskip

\begin{Thm}\label{thm1} Let $A,B,C,D$ be complex
matrices of size $n\times n$.

(i) For any positive integer $m$
\begin{equation} \label{p}
\int\limits_{U(n)} \hspace{-1ex}
{\det}^m[(AU\!+\!C)(BU\!+\!D)^{*}] dU = \int\limits_0^{\infty}
\hspace{-0.7ex}\ldots \hspace{-0.7ex} \int\limits_0^{\infty} \
 \prod\limits_{j=1}^m \det (CD^{*} \!+\! t_jAB^{*})\  d\mu_n(t_1, \ldots,
t_m)
\end{equation}

(ii) If $AA^{*}< CC^{*}$ and $BB^{*}< DD^{*}$ then for any
positive integer $m$ such that $2m \le n$
\begin{equation}\label{n}
 \int\limits_{U(n)} \hspace{-1ex} \frac{
dU}{{\det}^m [(AU\!+\!C)(BU\!+\!D)^{*}]}  = \int\limits_0^1
\hspace{-0.7ex}\ldots \hspace{-0.7ex} \int\limits_0^1
 \frac{d\nu_n(t_1, \ldots, t_m)}{\prod\limits_{j=1}^m \det
(CD^{*}\!-\!t_jAB^{*})}.
\end{equation}

\end{Thm}

\bigskip

\emph{Remark.} Identities (\ref{p}) -- (\ref{n}) may be written in
yet another form by making use of the well-known identity
\[
\int \ldots \int \det \left(p_j(t_i)\right)_{i,j=1}^m\ \det
\left(q_j(t_i)\right)_{i,j=1}^m \ dt_1\ldots dt_m=m! \ \det \left(
\int p_i(t)q_j(t)\ dt \right)_{i,j=1}^m.
\]
We have
\begin{equation}\label{p1}
\int\limits_{U(n)} {\det}^m[(AU+C)(BU+D)^{*}] dU = \frac{m!}{c_n}\
\det \left( \int\limits_0^{+\infty} \frac{\det(CD^{*}+tAB^{*})\
t^{i+j}\ dt }{(1+t)^{n+2m}}\right)_{i,j=0}^{m-1}
\end{equation}
and
\begin{equation}\label{n1}
\int\limits_{U(n)} \frac{ dU}{{\det}^m [(AU+C)(BU+D)^{*}]} =
\frac{m!}{k_n}\ \det \left( \int\limits_0^{1} \frac{(1-t)^{n-2m} \
t^{i+j}\ dt
 }{\det(CD^{*}-tAB^{*})}\right)_{i,j=0}^{m-1}.
\end{equation}

\bigskip

Obviously, by letting $C=D=zI$ in (\ref{p}) and (\ref{n}) one
obtains formulas for moments of the spectral determinants $|\det
(zI-AU)|^2$. In particular,
\begin{equation}\label{p2}
\int\limits_{U(n)} \hspace{-1ex} |\det (zI-AU)|^2 \ dU = (n+1)
\int\limits_0^{\infty} \frac{\det(|z|^2I+ t AA^*)}{(1+t)^{n+2}} \
dt
\end{equation}
and, provided $n\ge 2$,
\begin{equation}\label{n2}
\int\limits_{U(n)} \hspace{-1ex} \frac{ dU}{|\det (zI-AU)|^2}
=
\left\{%
\begin{array}{ll}
    \displaystyle{ \int\limits_0^1 \frac{(n-1)(1-t)^{n-2}}{\det (AA^*-t|z|^2I)} \
dt}, & \hbox{if $|z|^2 < \lambda_{min}(AA^*) $;} \\[2ex]
        \displaystyle{ \int\limits_0^1 \frac{(n-1)(1-t)^{n-2}}{\det (|z|^2I-tAA^*)} \
dt}, & \hbox{if $|z|^2>\lambda_{max}(AA^*)$;} \\
\end{array}%
\right.
\end{equation}
where $\lambda_{min}(AA^*)$ and $\lambda_{max}(AA^*)$ are
respectively the smallest and largest eigenvalues of $AA^*$.

If $\lambda_{min}(AA^*) \le |z|^2 \le \lambda_{max}(AA^*)$, then
the integral on the left-hand side in (\ref{n2}) should be handled
with care. One way to do this is to regularize the integrand.

For positive $\varepsilon$, define
\[
R_{z,\varepsilon} (A,A^*) = \int_{U(n)}
\frac{dU}{\det\left[\varepsilon^2I+\left(I-\frac{1}{z}AU\right)\left(I-\frac{1}{z}AU\right)^{*}\right]}.
\]
The integral on the right-hand side is, in fact, a function of
$AA^*$ and our next theorem evaluates this function in terms of
the eigenvalues of $AA^*$.

\bigskip

\begin{Thm}\label{thm2a} Let $\varepsilon >0$, and assume that $n\ge 2$.
Then for any $n\times n$ matrix $A$ and any non-zero complex $z$
\begin{equation*}
R_{z,\varepsilon} (A,A^*) =
 \frac{n-1}{2\pi
i} \int_0^1  (1-t)^{n-2}d t\ \int_{-\infty}^{+\infty}
\frac{dx}{x}\ \frac{1}{\det \left[\frac{1}{|z|^{2}} AA^*
+\left(\varepsilon^2 - t\right)I -i \varepsilon
\sqrt{t}\left(x+\frac{1}{x}\right)I\right]}.
\end{equation*}
If the eigenvalues $a_j^2$ of $AA^*$ are all distinct,
%$0\le\lambda_{min}(AA^*)=a_1^2< a_2^2 < \ldots <a_n^2=\lambda_{max}(AA^*)$,
then for any non-zero $z$ in the
annulus $\lambda_{min}(AA^*) < |z|^{2}< \lambda_{max}(AA^*)$ we
have
\begin{equation}\label{ef}
\lim_{\varepsilon \to 0} \frac{R_{z,\varepsilon} (A,A^*)}{\ln
(1/\varepsilon^2)} = (n-1)|z|^2 \sum_{j=1}^n \hspace{0.5ex}
(|z|^2- a_j^2)^{n-2} \hspace{0.5ex}\theta
 (|z|^2 -  a_j^2) \hspace{0.5ex} \prod_{k\not=j}\hspace{0.5ex}
 \frac{1}{a_k^2-a_j^2},
\end{equation}
where $\theta $ is the Heaviside step function, $\theta (r)=1/2$
at $r=0$.
\end {Thm}

\medskip

We prove Theorems \ref{thm1} and \ref{thm2a} in Sections
\ref{sec4} and \ref{sec5}, respectively, by making use of two
techniques, which to a certain extent are equivalent. One is based
on the expansion of moments of spectral determinants in characters
of the unitary group and subsequent use of the orthogonality of
characters. On this way, Theorem \ref{thm1} is equivalent to two
combinatorial identities (\ref{eq2:18}) and (\ref{eq2:19}), one of
which is a particular case of the Selberg integral in the form of
Kaneko \cite{Ka} and Kadell \cite{Kad}. We prove (\ref{eq2:18})
and (\ref{eq2:19}) in Section \ref{sec3}. These combinatorial
identities can be stated in the form of matrix integrals
(\ref{eq3:2a}) and (\ref{eq3:3a}) and are of independent interest.
They lead to evaluation of some non-trivial matrix integrals, as
discussed at the end of Section \ref{sec3}. The other technique is
based on the so-called color-flavor transformation, due to
Zirnbauer \cite{Z}. This transformation has many uses, and in the
random-matrix context it provides a very convenient tool to handle
moments of spectral determinants.

As an application of Theorem \ref{thm1}, let us consider random
matrices (\ref{con}). In the limit $n\to \infty$ the eigenvalues
of $W_n$ get closer and closer to the unit circle. Let ${\cal
N}_{n} (a,b)$ be the number of eigenvalues of $W_n$ in the annulus
\[
D_{a,b}=\left\{z: \frac{2a}{n}\le 1-|z|^2 \le
\frac{2b}{n}\right\}, \hspace{3ex} 0<a<b.
\]
By (\ref{int:eq2}),
\[
\langle {\cal N}_{n} (a,b) \rangle_{U(n)}= \int\limits_{D_{a,b}}
\left( \frac{n-1}{\pi \gamma |z|^2} \left(\frac{\tilde
\gamma}{\gamma} \right)^{n-2}\hspace{-2ex} \int\limits_{U(n-1)}
|\det (zI_{n-1} - \tilde G_{n-1}U_{n-1} )|^2 dU_{n-1}\right) dxdy.
\]
Making use of (\ref{p2}),
\[
\int\limits_{U(n-1)} |\det (zI_{n-1} - \tilde G_{n-1}U_{n-1} )|^2
dU_{n-1} = n \int_0^{\infty} \frac{[|z|^2+t
(1-\tilde\gamma)](|z|^2+t)^{n-2}}{(1+t)^{n+1}} \ dt
\]
and
\[
\frac{1}{n}\langle {\cal N}_{n} (a,b) \rangle_{U(n)} =\pi
\int_{1-\frac{2b}{n}}^{1-\frac{2a}{n}} f_n(q) dq,
\]
where
\[
f_n(q)= \frac{n-1}{\pi \gamma q} \left(\frac{\tilde
\gamma}{\gamma} \right)^{n-2} \int_0^{\infty} \frac{[q+t
(1-\tilde\gamma)](q+t)^{n-2}}{(1+t)^{n+1}} \ dt
\]
and $\tilde \gamma = (|z|^2+\gamma -1)/|z|^2$. Letting
$n\to\infty$, we obtain, after simple manipulations, that
\[
\lim_{n\to \infty} \frac{1}{n}\langle {\cal N}_{n} (a,b)
\rangle_{U(n)} = \frac{\sinh a}{a}
\exp\left(\frac{a(\gamma-2)}{\gamma}\right)
 - \frac{\sinh b}{b} \exp\left(\frac{b(\gamma-2)}{\gamma}\right),
\]
recovering one of the formulas of \cite{FSom2}, who, using a
different method requiring knowledge of the joint probability
distribution of eigenvalues, found the mean density of eigenvalues
and higher order correlation functions for the general case of
finite-rank deviation from the CUE. Note that when $\gamma =1$ the
nonzero eigenvalues of $G_nU_n$ coincide with the $(n-1)\times
(n-1)$ matrix obtained from $U_n$ by removing its first row and
column, see \cite{ZS} for more information about eigenvalue
statistics of truncated unitary matrices.

Now, we would like to elaborate on the point made at the end of
Introduction. Consider random complex matrices $W$ of the size
$n\times n$ with unitary invariant matrix distribution. Then, by
making use of the unitary invariance and Theorem \ref{thm1},
\begin{equation}\label{eq2:40}
\left\langle |\det (zI-W)|^2 \right\rangle_W = \left\langle
\int_{U(n)} |\det (zI-WU)|^2 dU \right\rangle_W = \int_0^{\infty}
\frac{p_n(|z|^2t)}{(1+t)^{n+2}}\ dt,
\end{equation}
where
\[
p_n(x)=\langle \det (xI+WW^*) \rangle_W.
\]
A similar formula holds for higher order moments of $|\det
(zI-W)|^2$. Thus, Theorem \ref{thm1} reduces the original
non-Hermitian problem to a Hermitian one.

The integral on the right-hand side in (\ref{eq2:40}) can be
evaluated, in the limit of infinite matrix dimension, in terms of
the limiting eigenvalue distribution of the non-negative matrices
$WW^*$. To this end, consider, for example, the complex $n\times
n$ matrices $W$ with the matrix distribution characterized by the
Feinberg-Zee density
\begin{equation}\label{fz}
Const. \times e^{-n\tr V(WW^*)},
\end{equation}
where $V(r)$ is a polynomial in $r$, $V(r)=a_mr^m + \ldots$,
$a_m>0$. Then
\[
p_n(x)=e^{n\int\ln (x+\lambda) dw(\lambda)}(1+o(1)),
\]
where $dw(\lambda)$ is the limiting normalised eigenvalue counting
measure of $WW^*$, and it can be shown that
\[
\lim_{n\to\infty}\frac{1}{n}\ln \left\langle |\det (zI-W)|^2
\right\rangle_W = \Phi (x,y),
\]
where
\begin{equation}\label{Phi}
\Phi (x,y) =
  \begin{cases}
    \ln |z|^2 & \text{if $|z| >m_1=\int \lambda dw (\lambda)$}, \\[1.5ex]
    \displaystyle{\int_0^{\infty} \ln \lambda dw (\lambda)} & \text{if $\displaystyle{\frac{1}{|z|} > m_{-1}=\int \frac{
dw (\lambda)}{\lambda}}$},\\[1.5ex]
     \displaystyle{|z|^2 + \int_0^{\infty} \ln \frac{\lambda +t_0}{|z|^2 +t_0} dw (\lambda)} & \text{if $\displaystyle{1/m_{-1}<|z| <m_1},$}
  \end{cases}
\end{equation}
where $t_0$ is the unique non-negative solution of
\[
\int_0^{\infty} \frac{dw(\lambda)}{\lambda +t}=\frac{1}{|z|^2 +
t}.
\]
The function $\Phi (x,y)$ is subharmonic and, hence, defines a
measure $d\nu = \frac{1}{4\pi}\Delta \Phi$ in the complex plane.
Here $\Delta$ is the Laplacian in variables $x$ and $y$. For the
Ginibre ensemble of random matrices this measure can be found
explicitly. In this case $V(r)=r$ and $WW^*$ is a Wishart ensemble
of random matrices. Its limiting eigenvalue distribution
$dw(\lambda)$ is given by $dw(\lambda)=
\frac{1}{2\pi}\sqrt{(4-\lambda)/\lambda}, \hspace{2ex} 0< \lambda
<4$, with $m_1=1$ and $m_{-1}=\infty$. A straightforward but
tedious calculation shows that $\Phi (x,y) = |z|^2-1$ inside the
unit disk $|z|^2<1$. Therefore $d\nu$ is the uniform distribution
on the unit disk, which is the same as the limiting eigenvalue
distribution in the Ginibre ensemble of random matrices, and,
hence, for this ensemble
\begin{equation}\label{ber}
\lim_{n\to\infty}\frac{1}{n}\left\langle\ln  |\det (zI-W)|^2
\right\rangle_W = \lim_{n\to\infty}\frac{1}{n}\ln
\left\langle|\det (zI-W)|^2 \right\rangle_W,
\end{equation}
so that the operations of taking logarithm and taking average
commute in the limit $n\to\infty$. A similar relation is known to
hold for Wigner ensembles of Hermitian matrices \cite{Berezin2}.

It would be interesting to investigate conditions on random matrix
distributions which guarantee (\ref{ber}). As the left hand-side
in (\ref{ber}) is the logarithmic potential of the limiting
eigenvalue distribution of $W$, this together with our Theorem
\ref{thm1} would give a useful tool for calculating eigenvalue
distributions in the complex plane. There are indications that the
range of matrix distributions for which (\ref{ber}) holds is quite
wide and contains the invariant ensembles (\ref{fz}). Indeed,
$\Phi (x,y)$ of equation (\ref{Phi}) reproduces the density of
eigenvalue distribution in ensembles (\ref{fz}) which was obtained
in \cite{FZ,FSZ} with the help of the method of
Hermitization\footnote{This method has a hidden regularization
procedure which has to justified to satisfy the mathematical
rigor.}.  In this context we would like to mention calculation of
Brown's measure for $R$-diagonal elements in finite von Neumann
algebras \cite{HL}, see also \cite{BL}. A matrix model for such
elements is provided by random matrices $RU$ where $U$ is random
unitary and $R$ is positive-definite, and Brown's measure is in a
way a regularized version of the eigenvalue distribution. Again,
$\Phi (x,y)$ of equation (\ref{Phi}) reproduces Brown's measure
found in \cite{HL}.

\section{Combinatorial identities} \label{sec3}

%Our proof of Theorems \ref{thm1} and \ref{thm2a} makes use of two
%combinatorial identities for Schur functions which we state and
%prove in this section.

\noindent {\bf Schur functions.}  \hspace*{1ex} In order to make
our paper self-contained we recall below the required facts from
the theory of symmetric polynomials.

A partition is a finite sequence $\lambda=(\lambda_1, \lambda_2,
\ldots , \lambda_n)$ of integers, called parts, such that
$\lambda_1 \ge \lambda_2 \ge \ldots \ge \lambda_n \ge 0$. The
weight of a partition, $|\lambda|$, is the sum of its parts
$|\lambda|=\sum_j \lambda_j$, and the length, $l(\lambda)$, is the
number of its non-zero parts. No distinction is made between
partitions which differ merely by the number of zero parts, and
different partitions of weight $r$ represent different ways to
write $r$ as a sum of natural numbers.

Partitions can be viewed as Young diagrams. The Young diagram of
$\lambda$ is a rectangular array of boxes (or dots), with
$\lambda_j$ boxes in the $j$th row, the rows being lined up on the
left. By transposing the diagram of $\lambda$ (i.e. interchanging
the rows and columns) one obtains another partition. This
partition is called conjugate of $\lambda$ and denoted by
$\lambda^{\prime}$. For example the conjugate of the partition
$(r)$ of length one is the partition $(1, \ldots , 1)\equiv (1^r)$
of length $r$. Obviously, $l(\lambda^{\prime})=\lambda_1$ and
$|\lambda|=|\lambda^{\prime}|$.

For any partition $\lambda$ of length $l(\lambda) \le n$
\begin{equation}\label{eq2:1}
s_{\lambda}(x_1, \ldots , x_n)=\frac{\det \left(
x_i^{\lambda_j+n-j}\right)_{i,j=1}^n}{\det\left(
x_i^{n-j}\right)_{i,j=1}^n}
\end{equation}
is a symmetric polynomial in $x_1, \ldots , x_n$, homogeneous of
degree $|\lambda|$. These polynomials are known as the Schur
functions. By convention, $s_{\lambda}(x_1, \ldots , x_n)=0$ if
$l(\lambda) >n$. This convention is in agreement with the apparent
identities
\begin{eqnarray}\label{eq2:1a}
s_{\lambda}(x_1, \ldots , x_{n-1}, 0) &=& s_{\lambda}(x_1, \ldots
, x_{n-1}) \hspace{3ex} \hbox{if $l(\lambda)\le n-1$}\\
    &=& \label{eq2:1b} 0 \phantom{s_{\lambda}(x_1, \ldots , x_{n-1})} \hspace{2ex}
\hbox{if $l(\lambda) >  n-1$.}
\end{eqnarray}

For partitions of length one, $\lambda=(r)$, the Schur functions
$s_{\lambda}$ are the complete symmetric functions $h_r$,
\begin{equation}\label{eq2:2}
s_{(r)}(x_1, \ldots, x_n)=h_r (x_1, \ldots, x_n)= \sum_{1\le i_1
\le i_2 \le \ldots \le i_r \le n} x_{i_1}x_{i_2} \ldots  x_{i_r},
\end{equation}
and $s_{\lambda^{\prime}}$ are the elementary symmetric functions
$e_r$,
\begin{equation}\label{eq2:3}
s_{(1^r)}(x_1, \ldots, x_n)=e_r (x_1, \ldots, x_n)= \sum_{1\le i_1
< i_2 < \ldots < i_r \le n} x_{i_1}x_{i_2} \ldots  x_{i_r}.
\end{equation}
More generally, see e.g. \cite{Macdonald} p. 41, the Jacobi-Trudi
identity asserts that for any $n \ge l(\lambda)$,
\begin{equation}\label{eq2:4}
s_{\lambda}=\det\left(h_{\lambda_i-i+j}\right)_{i,j=1}^n,
\hspace{2ex} s_{\lambda^{\prime}}=\det\left(e_{\lambda_i-i+j}
\right)_{i,j=1}^n
\end{equation}
where, by convention, $e_r=h_r=0$ if $r<0$.

We shall also need the Schur functions of matrix argument. If $M$
is an $n\times n$ matrix then
\[
s_{\lambda} (M) = s_{\lambda} (x_1, \ldots, x_n)
\]
where $x_1, \ldots, x_n$ are the eigenvalues of $M$. Thus
$s_{\lambda} (M)$ is a symmetric polynomial in the eigenvalues of
$M$. In view of (\ref{eq2:4}), it is also a polynomial in the
matrix entries of $M$. The Schur functions of matrix argument are
the characters of irreducible representations of the general
linear group and its unitary subgroup and, as a consequence, have
an important property of orthogonality. If $\lambda $ and $\mu$
are two partitions and $A$ and $B$ are two $n\times n$ matrices
then, see e.g. \cite{Macdonald} p. 445,
\begin{equation}\label{eq2:7}
\int_{U(n)} s_{\lambda} (AU) \overline{s_{\mu} (BU)} dU
=\delta_{\lambda,\mu}\frac{s_{\lambda}(AB^*)}{d_{\lambda}},
\end{equation}
and
\begin{equation}\label{eq2:7a}
\int_{U(n)} s_{\lambda} (AUBU^*) dU
=\frac{s_{\lambda}(A)s_{\lambda}(B)}{d_{\lambda}},
\end{equation}
where $d_{\lambda}$ is the dimension of the irreducible
representations of $U(n)$ with signature $\lambda$,
\[
d_{\lambda}= s_{\lambda}(I_n) = s_{\lambda}(1_n).
\]
We use the notation $(1_n)$ for the $n$-tuple $(1, \ldots, 1)$.

If $\lambda$ is a partition of length 1, $\lambda=(r)$,  then
\begin{equation}\label{eq2:5}
s_{\lambda}(1_n)=h_r (1_n)=\left(%
\begin{array}{c}
  n + r - 1 \\
  r \\
\end{array}%
\right)=\frac{(n+r-1)!}{r!(n-1)!},
\end{equation}
and
\begin{equation}\label{eq2:6}
s_{\lambda^{\prime}}(1_n)=e_r (1_n)=\left(%
\begin{array}{c}
  n \\
  r \\
\end{array}%
\right)=\frac{n!}{r!(n-r)!}.
\end{equation}
In general, explicit expressions are known for $s_{\lambda}(1_n)$
and $s_{\lambda^{\prime}}(1_n)$ in terms of the $\lambda_j$'s. If
$l(\lambda)\le n$ then for any $m\ge l(\lambda)$
\begin{equation}\label{eq2:8}
s_{\lambda}(1_n)=\left\{\prod_{1\le i <j\le
m}(\lambda_i-i-\lambda_j+j)\right\}\times \prod_{j=1}^m
\frac{(n+\lambda_j-j)!}{(m+\lambda_j-j)!(n-j)!}.
\end{equation}
If $l(\lambda^{\prime})\le n$ then for any $m\ge l(\lambda)$
\begin{equation}\label{eq2:9}
s_{\lambda^{\prime}}(1_n)=\left\{\prod_{1\le i <j\le
m}(\lambda_i-i-\lambda_j+j)\right\}\times \prod_{j=1}^m
\frac{(n+j-1)!}{(n+j-1-\lambda_j)!(m+\lambda_j-j)!}.
\end{equation}
%Identity (\ref{eq2:8}) is well known in the representation theory.
%We presume that identity (\ref{eq2:9}) is also known, however we
%have not found it in the literature.
Both identities can be derived by evaluating the binomial
determinants in (\ref{eq2:4}).

We shall also need the Cauchy identities for Schur functions, see,
e.g., \cite{Macdonald}, pp. 63, 65. Let $X$ be an $n\times n$
matrix. Then
\begin{eqnarray}\label{eq2:12}
 \prod_{i=1}^m \det(I_n+t_iX)&=&\sum_{\lambda}
s_{\lambda}(t_1, \ldots, t_m) s_{\lambda^{\prime}}(X)
 \\ \label{eq2:13}
 \prod_{i=1}^m \frac{1}{\det (I_n-t_iX)}&=&\sum_{\lambda}
s_{\lambda}(t_1, \ldots, t_m) s_{\lambda}(X).
\end{eqnarray}
The summation in (\ref{eq2:12}) is over all partitions such that
$l(\lambda) \le m$ and $l(\lambda^{\prime}) \le n$ and is finite.
The summation in (\ref{eq2:13}) is over all partitions such that
$l(\lambda) \le \min (m,n)$ and is infinite. The corresponding
series converges absolutely if $XX^*<I_n$.

\medskip

\noindent {\bf Beta-function determinants.}  \hspace*{1ex} When
$m=1$ identities (\ref{eq2:12}) and (\ref{eq2:13}) take the
familiar form of the expansion of the characteristic polynomial
and its reciprocal  in terms of the elementary symmetric functions
and complete symmetric functions, respectively. In this case
Theorem \ref{thm1} is a straightforward consequence of the
orthogonality property of the Schur functions (\ref{eq2:7}) and
the Euler integral
\begin{equation}\label{euler}
\int_0^{1} t^{p-1}(1-t)^{q-1}\, dt =
\int_0^{+\infty} \frac{t^{p-1}}{(1+t)^{p+q}}\
dt=\frac{\Gamma(p)\Gamma(q)}{\Gamma(p+q)}=B(p,q), \hspace{3ex} \re
p,q>0,
\end{equation}
where $\Gamma (p)$ and $B(p,q)$ are the Gamma and Beta functions
respectively. Indeed, for example,
\begin{eqnarray*}%\label{eq2:16}
\int_{U(n)}{\det}(I+AU)\det(I+BU)^*\ dU &=&\sum_{r=0}^n
\frac{e_r(AB^*)}{e_r(1_n)}\\ &=& (n+1) \int_0^{\infty} \det (I +
tAB^*)\ \frac{dt}{ (1+t)^{n+2} },
\end{eqnarray*}
where we have used (\ref{euler}) in the form
\begin{equation}\label{e}
\frac{1}{e_r(1_n)} =(n+1)\int_0^{+\infty}
\frac{t^{r}}{(1+t)^{n+2}}\ dt.
\end{equation}
Similarly,
\begin{equation}\label{h}
\frac{1}{h_r(1_n)} =(n-1) \int_0^{1} t^{r}(1-t)^{n-2}\, dt,
\end{equation}
and this identity does the trick for the reciprocal characteristic
polynomials.

Our proof of Theorem \ref{thm1} uses the following generalization
of (\ref{e}) -- (\ref{h}) to multivariate setting.

\bigskip

\begin{Lemma}\label{Lemma1} Let $m$ and $n$ be nonnegative integers.
\begin{itemize} \item[(a)]
For any partition $\lambda$ such that $l(\lambda)\le m$ and
$l(\lambda^{^\prime})\le n$
\begin{equation}\label{eq2:18}
\frac{s^2_{\lambda}(1_m)}{s_{\lambda^{\prime}}(1_n)}=\frac{1}{c_n}
\int\limits_0^{\infty} \hspace{-0.7ex}\ldots \hspace{-0.7ex}
\int\limits_0^{\infty} s_{\lambda}(t_1, \ldots t_m) \Delta^2 (t_1,
\dots, t_m) \prod_{j=1}^m \frac{dt_j}{(1+t_j)^{n+2m}}.
\end{equation}
\item[(b)]
If $2m\le n$ then for any partition $\lambda$ such that
$l(\lambda)\le m$
\begin{equation}\label{eq2:19}
\frac{s^2_{\lambda}(1_m)}{s_{\lambda}(1_n)}=\frac{1}{k_n}\int\limits_0^{1}
\hspace{-0.7ex}\ldots \hspace{-0.7ex} \int\limits_0^{1}
s_{\lambda}(t_1, \ldots t_m) \Delta^2 (t_1, \dots, t_m)
\prod_{j=1}^m (1-t_j)^{n-2m}dt_j.
\end{equation}
\end{itemize}
The normalization constants $c_n$ and $k_n$ are given in
(\ref{cnkn}), and $\Delta (t_1, \ldots, t_m)$ is the Vandermonde
determinant (\ref{vander}).
\end{Lemma}

\medskip

\emph{Remark.} Identity (\ref{eq2:19}) can be inferred from a
generalization of the Selberg Integral due to Kaneko \cite{Ka} and
Kadell \cite{Kad}, of which (\ref{eq2:19}) is a particular case
corresponding to a special choice of parameters. However, we are
not aware about any generalization of the Selberg Integral leading
to (\ref{eq2:18}). Below, we give an elementary proof of
(\ref{eq2:18}) and (\ref{eq2:19}) which is based on evaluating a
determinant consisting of Beta functions, see Proposition
\ref{Prop1} below. Our proof has a limited scope and does not
extend to the generality of Kaneko and Kadell formulas.

\bigskip

\emph{Proof.} Let $f_j=m+\lambda_j-j$, $j=1, 2, \ldots, m.$ If
$l(\lambda)\le m\le n$ and $l(\lambda^{^\prime})\le n$ then by
(\ref{eq2:8}) -- (\ref{eq2:9})
\[
\frac{s^2_{\lambda}(1_m)}{s_{\lambda^{\prime}}(1_n)}=\Delta(f_1,
\ldots, f_m) \times \left(\prod_{j=0}^{m-1}
\frac{1}{j!^2(n+j)!}\right)\times \left(\prod_{j=1}^m
f_j!(n+m-1-f_j)!\right),
\]
where
\[
\Delta(f_1, \ldots, f_m) = \prod_{1\le i<j\le m} (f_i-f_j) = \det
\left(f_j^{m-i}\right)_{i,j=1}^m.
\]
By adding rows in the Vandermonde determinant $\det
\left(f_j^{m-i}\right)_{1}^m$,
\[
\Delta(f_1, \ldots, f_m) =
\det\left(p_{m-i}(f_j)\right)_{i,j=1}^m,
\]
where
\[
p_k(x)=(x+1)(x+2) \ldots (x+k).
\]
Hence
\begin{equation}\label{eq2:20}
f_1! f_2! \dots f_m! \Delta(f_1, \ldots, f_m)= \det \left(
(f_j+m-i)!\right)_{i,j=1}^{m},
\end{equation}
and
\[
\frac{s^2_{\lambda}(1_m)}{s_{\lambda^{\prime}}(1_n)}=
\left(\prod_{j=0}^{m-1} \frac{(n+m+j)!}{(j!)^2(n+j)!}\right)
\times \det \left(\ B(f_j+m-i+1, n+m-f_j)\ \right)_{i,j=1}^{m},
\]
where $B$ is the Beta function. By making use of Proposition
\ref{Prop1} below,
\begin{eqnarray*}
\det \left(\ B(f_j+\!m\!-\!i\!+\!1, n\!+\!m\!-\!f_j)\
\right)_{i,j=1}^{m}\!\!&\!\!=\!\!&\!\!\det \left(\
B(f_j+\!m\!-\!i\!+\!1, n\!+\!m\!-\!f_j+\!i\!-\!1)\
\right)_{i,j=1}^{m}\\ \!\!&\!\!=\!\!&\!\! \det \left(\
\int\limits_0^{+\infty} \frac{t^{f_j}t^{m-i}dt}{(1+t)^{n+2m}}\
\right)_{i,j=1}^m.
\end{eqnarray*}
It is apparent that
\begin{eqnarray*}
\det \left(\!  \int\limits_0^{+\infty}
\frac{t^{f_j}t^{m-i}dt}{(1+t)^{n+2m}}\!
\right)_{i,j=1}^m\!\!&\!\!=\!\!&\!\! \int\limits_0^{+\infty}
\hspace{-0.7ex}\cdots \hspace{-0.7ex} \int\limits_0^{+\infty}
\hspace{-0.7ex} s_{\lambda}(t_1, \ldots t_m) \det \left(
t_{i}^{m-j}\right)_{i,j=1}^{m}\ \prod_{i=1}^m
\frac{t_i^{m-i}dt_i}{(1+t_i)^{n+2m}}\\
\!\!&\!\!=\!\!&\!\!\frac{1}{m!} \int\limits_0^{+\infty}
\hspace{-0.7ex}\cdots \hspace{-0.7ex} \int\limits_0^{+\infty}
\hspace{-0.7ex} s_{\lambda}(t_1, \ldots t_m) \left[\det \left(
t_{i}^{m-j}\right)_{i,j=1}^{m}\right]^2 \prod_{i=1}^m
\frac{dt_i}{(1+t_i)^{n+2m}},
\end{eqnarray*}
and (\ref{eq2:18}) follows.

Similarly, if $2m\le n$ and $l(\lambda) \le m$ then by
(\ref{eq2:8})
\[
\frac{s^2_{\lambda}(1_m)}{s_{\lambda}(1_n)}=\Delta(f_1, \ldots ,
f_m) \times \left(\prod_{j=0}^{m-1} \frac{(n-j-1)!}{j!^2}\right)
\times \left(\prod_{j=1}^m \frac{f_j!}{(n-m+f_j)!}\right),
\]
and, in view of (\ref{eq2:20}),
\begin{equation}\label{extra1}
\frac{s^2_{\lambda}(1_m)}{s_{\lambda}(1_n)}= \left(
\prod_{j=0}^{m-1} \frac{(n-j-1)!}{(j!)^2(n-m-j-1)!} \right)\times
\det \left(\ B(f_j+\!m\!-\!i\!+\!1,n\!-\!2m\!+\!i)\
\right)_{i,j=1}^m.
\end{equation}
By Proposition \ref{Prop1},
\begin{eqnarray} \label{extra2}
\det \left( B(f_j+\!m\!-\!i\!+\!1,n\!-\!2m\!+\!i)
\right)_{i,j=1}^m\!\!&\!\!=\!\!&\!\!\det \left(
B(f_j+\!m\!-\!i\!+\!1,n\!-\!2m\!+\!1) \right)_{i,j=1}^m \\
\label{extra3} \!\!&\!\!=\!\!& \!\!\det \left( \int_0^1
t^{f_j}t^{m\!-\!i}(1\!-\!t)^{n-2m}\ dt \right)_{i,j=1}^m,
\end{eqnarray}
and (\ref{eq2:19}) follows. \hfill $\Box$

\bigskip

\begin{Prop}\label{Prop1} For any $p_1, p_2, \ldots , p_m$ and
$q_1, q_2, \ldots , q_m$ such that $\re p_j >m$ and $\re q_j
> -1$ we have
\begin{equation}\label{eq2:21}
\det \left(\ B(p_j-i, q_j+i)\ \right)_{1}^m = \det \left(\
B(p_j-i, q_j+1)\ \right)_{1}^m.
\end{equation}
\end{Prop}

\smallskip

\emph{Proof.} We shall use the identity
\begin{equation}\label{eq2:22}
B(p,q-1)+B(p-1, q)= B(p-1,q-1)
\end{equation}
and the operation of addition of columns to transform the
determinant on the left in (\ref{eq2:21}) to the one on the right.

It is convenient to write determinants by showing their columns.
With this convention,
\begin{equation}\label{eq2:23}
|B(p-1,q+1), B(p-2,q+2), \ldots, B(p-m+1,q+m-1), B(p-m,q+m)|
\end{equation}
represent the determinant on the l.h.s. in (\ref{eq2:21}). Let us
label its columns by numbers $1, \ldots, m$ from left to right (so
that the leftmost column is column $1$). Note a particular
property of columns in this determinant. As we move from column
$j$ to column $j+1$ the first argument of the Beta function
decreases by one 1 the second argument increases by 1. To be able
to refer to this property, we say that columns $1,2, \ldots, m$
are balanced.

Observing that column 1 has the desired form already, let us
perform the following operation on columns $2, 3, \ldots , m$.
Starting at column $m$ and working backwards, let us add to each
column the one that precedes it. In view of (\ref{eq2:22}) and the
above mentioned property of columns, this operation yields
\[
|B(p-1,q+1), B(p-2,q+1), B(p-3, q+2), \ldots, B(p-m+1,q+m-2),
B(p-m,q+m-1)|.
\]
Observing that columns 1 and 2 have the desired form now, and that
columns $2, \ldots , m$ remain balanced, we apply our operation
again, now on columns $3, \ldots, m$. This yields the determinant
\[
|B(p-1,q+1), B(p-2,q+1), B(p-3, q+1), \ldots, B(p-m+1,q+m-3),
B(p-m,q+m-2)|,
\]
where columns $1, 2, 3$ have the desired form and columns $4,
\ldots, m$ are balanced. It is clear that repeated application of
our operation will yield the determinant
\[
|B(p-1,q+1), B(p-2,q+1), B(p-3, q+1), \ldots, B(p-m,q+1)|
\]
after the final step. This is exactly the determinant on the right
in (\ref{eq2:22}). \hfill $\Box$

\medskip

\noindent {\bf Applications to matrix integrals.}  \hspace*{1ex}
In view of integration formulas (\ref{tt}) and (\ref{ss}),
identities (\ref{eq2:18}) and (\ref{eq2:19}) can be rewritten as:
\begin{equation}\label{eq3:2}
\int_{ZZ^*\ge 0} s_{\lambda}(ZZ^*)\ d\hat \mu_n (Z) =
\frac{s^2_{\lambda}(I_m)}{s_{\lambda^{\prime}}(I_n)}, \hspace{3ex}
\int_{ZZ^*\le I_m} s_{\lambda}(ZZ^*)\ d\hat \nu_n (Z) =
\frac{s^2_{\lambda}(I_m)}{s_{\lambda}(I_n)},
\end{equation}
where $Z$ are complex $m\times m$ matrices, and $I_m$ and $I_n$
are identity matrices of sizes $m\times m$ and $n\times n$,
respectively. The first identity holds for any non-negative
integer $n$ and any partition $\lambda$ such that
$l(\lambda^{\prime})\le n$. The second one holds for any integer
$n\ge 2m$ and any $\lambda$. Since the above identities become
trivial (both sides vanish) for partitions of length $>m$ we drop
the restriction $l(\lambda) \le m$.

These two identities lead to several useful matrix integrals.

Let $M$ be an $m\times m$ matrix. Then, for any non-negative
integer $n$,
\begin{equation}\label{eq3:2a}
\int_{ZZ^*\ge 0} s_{\lambda}(MZZ^*)\ d\hat \mu_n (Z) =
\frac{s_{\lambda} (M) s_{\lambda}(I_m)}{s_{\lambda^{\prime}}(I_n)}
\end{equation}
provided $l(\lambda^{\prime})\le n$, and if  $n\ge 2m$ then
\begin{equation}\label{eq3:3a}
\int_{ZZ^*\le I_m} s_{\lambda}(MZZ^*)\ d\hat\nu_n (Z) =
\frac{s_{\lambda} (M) s_{\lambda}(I_m)}{s_{\lambda}(I_n)}
\end{equation}
for any $\lambda$. These two integrals follow from (\ref{eq3:2}) and
(\ref{eq2:7a}) and the unitary invariance of  $d\hat \mu_n (Z)$ and
$d\hat\nu_n (Z)$.

If  $L$ and $M$ are two $m\times m$ matrices then for any
non-negative integer $n$
\begin{equation}\label{eq3:2aa}
\int_{ZZ^*\ge 0}s_{\lambda}(LZ)\overline{s_{\mu}(MZ)}\
d\hat\mu_n(Z)=\delta_{\lambda,\mu}\frac{s_{\lambda}(LM^*)}{s_{\lambda^{\prime}}(I_n)},
\end{equation}
provided $l(\lambda^{\prime}) \le n$ and $l(\mu^{\prime}) \le n$,
and if $n\ge 2m$ then
\begin{equation}\label{eq3:3aa}
\int_{ZZ^*\le I_m} s_{\lambda}(LZ)\overline{s_{\mu}(MZ)}\
d\hat\nu_n (Z) =
\delta_{\lambda,\mu}\frac{s_{\lambda}(LM^*)}{s_{\lambda} (I_n)},
\end{equation}
These orthogonality relations follow from (\ref{eq2:7}) and
(\ref{eq3:2a}) -- (\ref{eq3:3a}), and, in turn, lead to
Berezin-Hua integrals \cite{Hua,Berezin1}
\begin{eqnarray*}
\int_{ZZ^*\ge 0}  {\det}^n (I_m+LZ){\det}^n (I_m+MZ)^*\ d\hat
\mu_n (Z) &=& {\det}^n (I_m+LM^*)
\\
\int_{ZZ^*\le I_m}  \frac{d\hat \nu_n (Z)}{{\det}^n (I_m-LZ)
{\det}^n (I_m-MZ)^*} &=& \frac{1}{{\det}^n (I_m-LM^*)},
\hspace{1ex} n\ge 2m.
\end{eqnarray*}
One only has to recall the Cauchy identites (\ref{eq2:12}) and
(\ref{eq2:13}).

If $P$ and $Q$ are two $n\times m$ matrices and $n\ge 2m$ then it
follows from  (\ref{eq2:7}) and (\ref{eq3:3aa}) that
\begin{equation}\label{prez}
\int_{U(n)} s_{\lambda}(PQ^{*}U)\overline{s_{\mu}(PQ^{*}U)}\ dU =
\int_{ZZ^*\le I_m}
s_{\lambda}(P^{*}PZ)\overline{s_{\mu}(Q^{*}QZ)}\ d\hat\nu_n (Z)
\end{equation}
for any $\lambda$ and $\mu$. Identity (\ref{prez}) implies that
\begin{equation}\label{eq3:7}
\int_{U(n)} e^{\tr (PQ^{*}U + U^{*}QP^{*})}\ dU = \int_{ZZ^*\le
I_m} e^{\tr (P^{*}PZ + Z^{*}Q^{*}Q)} \ d\hat\nu_n(Z).
\end{equation}
The duality relation (\ref{eq3:7}) is a particular case of
Zirnbauer's color-flavor transformation \cite{Z}. It can be easily
obtained from (\ref{prez}) by making use of the expansion
\begin{equation}\label{eq3:8}
e^{\tr A}=\sum_{\lambda} c_{\lambda} s_{\lambda} (A).
\end{equation}
In fact, (\ref{eq3:7}) extends to any series $g(A)=\sum_{\lambda}
c_{\lambda} s_{\lambda} (A)$,
\[
\int_{U(n)} |g(PQ^{*}U)|^2\ dU = \int_{ZZ^*\le I_m} g(P^{*}PZ)
\overline{g(Q^{*}QZ)} \ d\hat\nu_n(Z).
\]

It follows from (\ref{eq3:8}) and  (\ref{eq2:7}) that the integral
over the unitary group on the left-hand side in (\ref{eq3:7}) is a
function of $Q^{*}QP^{*}P$. This function can be evaluated
explicitly in terms of the eigenvalues of $Q^{*}QP^{*}P$. We would
like to demonstrate this in a slightly more general setting.

For square matrices $A$ and $B$ of size $n\times n$ define
\begin{equation}\label{eq3:ad1}
F_n(AB^*)= \int_{U(n)} e^{\tr (AU + U^* B^*)}\ dU.
\end{equation}
If the eigenvalues $z_1^2, \ldots , z_n^2$ of the matrix $AB^*$
are all distinct then \cite{SW}
\[
F_n(AB^*) = \frac{Const.}{\Delta (z^2_1, \ldots , z^2_n)}   \times
\det \left( z_i^{j-1} I_{j-1}(z_i) \right)_{i,j=1}^n
\]
where $I_k$ is the modified Bessel function,
\[
I_k(z)=\sum_{j=0}^{\infty}
\frac{\left(\frac{z}{2}\right)^{2j+k}}{j!(j+k)!}.
\]
For our purposes, we want to know $F_n(AB^*)$ for matrices $AB^*$
of low rank, e.g. when $AB^*$ is rank one.

\bigskip

\begin{Lemma}\label{Lemma5} Suppose that $AB^*$ has $m$ distinct non-zero
eigenvalues $z^2_1, \ldots , z^2_m$ and $2m\le n$. Then
%\begin{equation}\label{eq3:12}
\[
F_n(AB^*)=\left(\prod_{j=1}^m\frac{(n-j)!}{(n-m-j)!}\right)
\int\limits_0^{1} \hspace{-0.7ex}\ldots \hspace{-0.7ex}
\int\limits_0^{1}\  \frac{\det \left( g(t_iz_j^2)
\right)_{i,j=1}^m}{\Delta(z_1^2, \ldots , z_m^2)}\ \prod_{i=1}^m
t_i^{m-i} (1-t_i)^{n-2m}dt_i,
\]
%\end{equation}
where $ g(x)= I_0 \left( 2\sqrt{x} \right). $ In particular, if
$AB^*$ is rank one and $z^2$ is its non-zero eigenvalue then
\begin{equation}\label{eq3:12a}
F_n(AB^*)= (n-1) \int_0^1 I_0 \left(2\sqrt{tz^2}\right)
(1-t)^{n-2}\ dt.
\end{equation}

\end{Lemma}

\smallskip

\emph {Proof.} It follows from (\ref{eq3:8}) and  (\ref{eq2:7})
that
\[
F_n(AB^*) = \sum_{\lambda}
 \frac{c^2_{\lambda}}{s_{\lambda}(1_n)}s_{\lambda} (AB^*).
\]
The coefficients $c_{\lambda}$ are given by
\[
c_{\lambda} = \det \left( \frac{1}{(\lambda_j-j+i)!}
\right)_{i,j=1}^{m} = s_{\lambda}(1_m)\  \prod_{j=1}^m
\frac{(m-j)!}{(m+\lambda_j-j)! }
\]
see, e.g.,  \cite{Balantekin} and references therein, and
\[
F_n(AB^*)= \left(\prod_{j=0}^{m-1} j!^2 \right) \sum_{\lambda}
\frac{s^2_{\lambda}(1_m)}{s_{\lambda}(1_n)}\ \frac{s_{\lambda}
(AB^*)}{f_1!^2 \cdot \ldots \cdot f_m!^2 },
\]
where as before $f_j=m+\lambda_j-j$. Note that the summation is
over all partitions $\lambda$ of length $\le m$, or, equivalently,
over all $f_1>f_2 > \ldots > f_m \ge 0$. It follows now from
(\ref{extra1}) -- (\ref{extra2}) that
\[
F_n(AB^*)= \left(\prod_{j=1}^m\frac{(n-j)!}{(n-m-j)!}\right)
\int\limits_0^{1} \hspace{-0.7ex}\ldots \hspace{-0.7ex}
\int\limits_0^{1} \frac{g(t_1, \ldots t_m)}{\Delta(z_1^2,\ldots,
z_m^2) }\ \prod_{i=1}^m t_i^{m-i} (1-t_i)^{n-2m}dt_i.
\]
where
\begin{equation*}
g(t_1, \ldots t_m) = \sum_{f_1 > f_2 > \ldots > f_m\ge 0}
\frac{\det\left( t_i^{f_j} \right)_{i,j=1}^m \ \det\left(
z_i^{2f_j}\right)_{i,j=1}^m}{f_1!^2 \ldots f_m!^2}.
\end{equation*}
To complete the proof, recall the following generalization of the
Cauchy-Binet formula, see e.g. \cite{Hua} p. 22. If
$g(x)=\sum_{f\ge 0} \gamma_f x^f$ is an analytic function in the
complex $x$-plane then
\[
\det \left( g(t_ix_j)\right)_{i,j=1}^m = \sum_{f_1 > f_2
> \ldots > f_m\ge 0} \gamma_{f_1} \ldots \gamma_{f_m}\  \det \left(
t_i^{f_j}\right)_{i,j=1}^m  \  \det \left( x_i^{f_j}
\right)_{i,j=1}^m.
\]
By making use of this formula,
\[
g(t_1, \ldots, t_m) =\det \left(
I_{0}\left(2\sqrt{t_iz_j^2}\right) \right)_{i,j=1}^m,
\]
and Lemma follows. \hfill $\Box$

\bigskip

\section{Proof of Theorem \ref{thm1}} \label{sec4}

After all the preparatory work of the previous section, Theorem
\ref{thm1} becomes almost evident.

We first prove (\ref{p}) -- (\ref{n}) for $C=D=I$. With Lemma
\ref{Lemma1} in hand, this becomes a routine calculation.
Expanding powers of determinants in the Schur functions as in
(\ref{eq2:12}) -- (\ref{eq2:13}) and integrating over the unitary
group with the help of (\ref{eq2:7}), one gets
\begin{equation}\label{eq2:30}
\int\limits_{U(n)} \hspace{-1ex} {\det}^m[(I + AU)(I+ BU)^{*}] dU
= \sum_{\lambda}
\frac{s_{\lambda}^2(1_m)}{s_{\lambda^{\prime}}(1_n)}\,
s_{\lambda^{\prime}} (AB^*)
\end{equation}
and
\begin{equation}\label{eq2:31}
\int\limits_{U(n)} \hspace{-1ex} \frac{dU}{{\det}^m[(I  - AU)(I -
BU)^{*}]} = \sum_{\lambda}
\frac{s_{\lambda}^2(1_m)}{s_{\lambda}(1_n)}\, s_{\lambda} (AB^*).
\end{equation}
The sum in (\ref{eq2:30}) is finite and the sum in (\ref{eq2:31})
is absolutely converging for any $AA^* < I$ and $BB^*<I$. Now, by
making use of (\ref{eq2:18}) and (\ref{eq2:19}), and then
(\ref{eq2:7}) again, one arrives at
\begin{equation} \label{pa}
\int\limits_{U(n)} \hspace{-1ex} {\det}^m[(I + AU)(I+ BU)^{*}] dU
= \int\limits_0^{\infty} \hspace{-0.7ex}\ldots \hspace{-0.7ex}
\int\limits_0^{\infty} \
 \prod\limits_{j=1}^m \det (I+ t_jAB^{*})\  d\mu_n(t_1, \ldots,
t_m)
\end{equation}
and
\begin{equation}\label{na}
\int\limits_{U(n)} \hspace{-1ex} \frac{dU}{{\det}^m[(I  - AU)(I -
BU)^{*}]}  = \int\limits_0^1 \hspace{-0.7ex}\ldots \hspace{-0.7ex}
\int\limits_0^1
 \frac{d\nu_n(t_1, \ldots, t_m)}{\prod\limits_{j=1}^m \det
(I- t_jAB^{*})}.
\end{equation}

Extending (\ref{pa}) and (\ref{na}) to the generality of (\ref{p})
and (\ref{n}) is straightforward. If $C$ and $D$ are not
degenerate, then
\[
\int\limits_{U(n)} \hspace{-1ex} {\det}^m[(AU+C)(BU+D)^{*}] dU =
{\det}^m (C D^{*})\hspace{-1ex}  \int\limits_{U(n)} \hspace{-1ex}
{\det}^m[(I+ C^{-1}AU)(I + D^{-1}BU)^{*}] dU
\]
and (\ref{p}) follows from (\ref{pa}). The assumption that $C$ and
$D$ are not degenerate can be removed by the continuity argument.
\hfill $\Box$

\section{Regularization of the inverse determinant} \label{sec5}

In this section we employ another approach to the problem of
evaluating of negative moments of spectral determinants which is
to write the determinants as Gaussian integrals and then perform
the integration over the unitary group with the help of the
color-flavor transformation. This approach is not new. It was
pioneered by Zirnbauer in the context of unitary random matrix
ensembles. The new element here is that we apply it in the general
context of complex matrices.

We shall write the spectral determinant $\det[(I-AU)(I-AU)^*]$ of
$n\times n$ matrices as $2n\times 2n$ block determinant
\[
\det[(I-AU)(I-AU)^*]=\det[(U^*-A)(U^*-A)^*]=
\left|\begin{array}{cc}
  0 & i(U^*-A) \\
  i(U^* -A)^* & 0
\end{array} \right|
\]
and more generally
\[
\det[\varepsilon^2 I + (I-AU)(I-AU)^*]=\left|\begin{array}{cc}
    \varepsilon I & i(U^*-A) \\
  i(U^* -A)^* & \varepsilon I
\end{array} \right|
\]

\medskip

\begin{Prop} \label{sec4:prop1}
Suppose that $\re \lambda_j >0$, $j=1,2$. Then for any complex
$n\times n$ matrix $\Omega$ we have
\begin{equation}\label{sec4:2}
\left|\begin{array}{cc}
  \lambda_1I & i\Omega \\
  i\Omega^* & \lambda_2I
\end{array} \right|^{-1} = \frac{1}{\pi^{n}}\int_{\C^n}\hspace{-1.5ex}\dd
\vec{v} \int_{\C^n}\hspace{-1.5ex} \dd \vec{w} \hspace{1ex} \exp
\left\{ {-[\lambda_1\vec{v}^*\vec{v} +\lambda_2\vec{w}^*\vec{w} +i
(\vec{w}^* \Omega^* \vec{v} + \vec{v}^* \Omega \vec{w})]} \right\}.
\end{equation}
The integral on the right-hand side converges absolutely.
\end{Prop}

\smallskip

{\em Remark.} In this section, we shall use letters in bold face to
represent column vectors in $\C^n$. The symbol $\dd \vec{v}$ will
denote the volume element of $\vec{v}$ in $\C^n$,
\[
\dd \vec{v} = \prod_{j=1}^n \dd \eta_j = \prod_{j=1}^n d \re v_j \ d
\im v_j.
\]

\smallskip

\emph{Proof.} Note that
\begin{equation*}
\lambda_1\vec{v}^*\vec{v} + \lambda_2\vec{w}^*\vec{w} +i (\vec{w}^*
\Omega^* \vec{v} + \vec{v}^* \Omega \vec{w}) = (\vec{v}^*,\vec{w}^*)
\begin{pmatrix}
  \lambda_1I & i\Omega  \\
  i\Omega^* & \lambda_2I
\end{pmatrix}
 \begin{pmatrix}
   \vec{v} \\
   \vec{w} \
 \end{pmatrix}.
\end{equation*}
In view of the singular value decomposition  $\Omega = U^*\omega V$,
\[
\begin{pmatrix}
  \lambda_1I & i\Omega  \\
  i\Omega^* & \lambda_2I
\end{pmatrix}=
\begin{pmatrix}
  U^* & 0 \\
  0 & V^*
\end{pmatrix}
\begin{pmatrix}
  \lambda_1I & i \omega  \\
  i \omega & \lambda_2I
\end{pmatrix}
\begin{pmatrix}
  U & 0  \\
  0 & V
\end{pmatrix},
\]
where $\omega$ is diagonal matrix of singular values of $\Omega$,
$\omega =\diag (\omega_1, \ldots, \omega_n)$, and $U$ and $V$ are
unitary matrices. Introducing $\vec{f}=U\vec{v}$ and
$\vec{g}=V\vec{w}$,
\begin{eqnarray*}
(\vec{v}^*,\vec{w}^*) \begin{pmatrix}
  \lambda_1I & i\Omega  \\
  i\Omega^* & \lambda_2I
\end{pmatrix}
 \begin{pmatrix}
   \vec{v} \\
   \vec{w} \
 \end{pmatrix}
&=& (\vec{f}^*, \vec{g}^*)
\begin{pmatrix}
  \lambda_1I & i\omega  \\
  i\omega & \lambda_2I
\end{pmatrix} \begin{pmatrix}
  \vec{f}   \\
  \vec{g}
\end{pmatrix}\\
 &=& \sum_{j=1}^n\  (\bar f_j, \bar g_j) \begin{pmatrix}
  \lambda_1 & i\omega_j  \\
  i\omega_j & \lambda_2
\end{pmatrix}
\begin{pmatrix}
  f_j  \\
  g_j
\end{pmatrix}.
\end{eqnarray*}
Since $U$ and $V$ are unitary, $\dd \vec{v} = \dd \vec{f}$ and $\dd
\vec{w} = \dd \vec{g}$. Changing the variables of integration in
(\ref{sec4:2}) from $\vec{v}$ and $\vec{w}$ to $\vec{f}$ and
$\vec{g}$ breaks this $2n$-fold integral into the product of the
2-fold integrals
\[
\frac{1}{\pi}\int_{\C^2}
 \exp
 \left[
    -\lambda_1|f_j|^2 - \lambda_2|g_j|^2
    -i\omega_j( f_j\bar g_j + \bar f_j g_j )
 \right]
\dd f_j \dd g_j = (\lambda_1\lambda_2+\omega_j^2)^{-1}.
\]
Thus, the integral on the right-hand side in (\ref{sec4:2}) equals $
\prod_{j=1}(\lambda_1\lambda_2+\omega_j^2)^{-1} $ which is obviously
same as the determinant on the left-hand side. \hfill $\Box$

\medskip

\emph{Proof of Theorem \ref{thm2a}.} Obviously, without loss of
generality we can put $z=1$. Let
\begin{equation}\label{eq4:0}
R_{\varepsilon} (A,A^*) =\int_{U(n)} \frac{dU}{\det[\varepsilon^2I
+ (I-AU)(I-AU)^*] }, \hspace{3ex} \varepsilon >0.
\end{equation}
The integral on the right-hand side converges for any $n\times n$
matrix $A$.

It follows from Proposition \ref{sec4:prop1} that
\begin{equation}\label{eq4:3}
R_{\varepsilon} (A,A^*) =
\frac{1}{\pi^{n}}\int_{\C^n}\hspace{-1.5ex}\dd \vec{v}
\int_{\C^n}\hspace{-1.5ex} \dd \vec{w} \hspace{1ex} e^{-[\varepsilon
(\vec{v}^*\vec{v} +\vec{w}^*\vec{w}) -i(\vec{w}^* A^* \vec{v} +
\vec{v}^* A \vec{w})]} f_n(\vec{v}^*\vec{v}\vec{w}^*\vec{w}),
\end{equation}
where, cf. (\ref{eq3:ad1}),
\[
f_n(\vec{v}^*\vec{v}\vec{w}^*\vec{w})= \int_{U(n)} e^{i(\vec{w}^*
U\vec{v} +\vec{v}^*U^*\vec{w})} dU= \int_{U(n)} e^{i\tr
(\vec{v}\vec{w}^* U+U^*\vec{w}\vec{v}^*)} dU.
\]
By Lemma \ref{Lemma5},
\begin{equation}\label{eq4:5}
f_n(\vec{v}^*\vec{v}\vec{w}^*\vec{w})= \int_0^1 J_0\left(2\sqrt{t\,
\vec{v}^*\vec{v}\, \vec{w}^*\vec{w} }\right)\, d\sigma_n(t),
\end{equation}
where
\[
d\sigma_n(t) = (n-1)(1-t)^{n-2} dt
\]
and $J_0$ is the Bessel function
\[
J_0(z)=\sum_{j=0}^{\infty}
\frac{\left(\frac{iz}{2}\right)^{2j}}{j!^2}=I_0(iz).
\]
We have $|f_n(\vec{v}^*\vec{v}\vec{w}^*\vec{w})|\le 1 $ for all
$\vec{v}$ and $\vec{w}$. This is because $|J_0(z)| \le 1 $ for all
$z$. Therefore we can interchange the order of integrations on
replacing $f_n$ in (\ref{eq4:3}) by the integral of (\ref{eq4:5}).
This yields
\begin{equation}\label{eq4:7}
R_{\varepsilon} (A,A^*) = \int_0^1 d\sigma_n(t)\
\frac{1}{\pi^{n}}\int_{\C^n}\hspace{-1.5ex}\dd \vec{v}
\int_{\C^n}\hspace{-1.5ex} \dd \vec{w} \hspace{1ex} e^{-\varepsilon
(\vec{v}^*\vec{v} +\vec{w}^*\vec{w}) +i(\vec{w}^* A^* \vec{v} +
\vec{v}^* A \vec{w})} J_0\left(2\sqrt{t\, \vec{v}^*\vec{v}\,
\vec{w}^*\vec{w} }\right).
\end{equation}
In order to perform the integration in variables $\vec{v}$ and
$\vec{w}$ we shall make use of the integral representation
\begin{equation}\label{eq4:6}
J_0(2\sqrt{p\,q})=\frac{1}{2\pi i}\int_{-\infty}^{+\infty}
\frac{dx}{x}\ e^{i(px+\frac{q}{x})}
\end{equation}
which holds any $p>0$ and $q>0$ and is a particular case of
equation 3.871.1 in \cite{GR}. The integral in (\ref{eq4:6})
converges because of the oscillations of the exponential function,
however the convergence is not absolute.

We have
\begin{equation}\label{eq4:8}
J_0\left(2\sqrt{t\, \vec{v}^*\vec{v}\, \vec{w}^*\vec{w} }\right)
=\frac{1}{2\pi i}\int_{-\infty}^{+\infty} \frac{dx}{x}\
e^{i\sqrt{t}(x\vec{v}^*\vec{v}+\frac{\vec{w}^*\vec{w}}{x})}.
\end{equation}
Note that by Proposition \ref{sec4:prop1},
\[
\frac{1}{\pi^{n}}\!\int_{\C^n}\hspace{-1.5ex}\dd
\vec{v}\hspace{-0.5ex} \int_{\C^n}\hspace{-1.5ex} \dd \vec{w}
\hspace{1ex} e^{-\varepsilon (\vec{v}^*\vec{v} +\vec{w}^*\vec{w})
+i(\vec{w}^* A^* \vec{v} + \vec{v}^* A \vec{w})} \
e^{i\sqrt{t}(x\vec{v}^*\vec{v}+\frac{\vec{w}^*\vec{w}}{x})} =
\left|\hspace{-1ex}\begin{array}{cc}
  (\varepsilon - i\sqrt{t}x)I & -A \\
  -A^* & (\varepsilon - i\frac{\sqrt{t}}{x})I
\end{array} \hspace{-1ex} \right|^{-1}.
\]
Therefore, on replacing $J_0$ in (\ref{eq4:7}) by the integral of
(\ref{eq4:8}) and changing the order of integrations one arrives
at
\begin{equation}\label{eq4:9}
 R_{\varepsilon} (A,A^*) = \int_0^1 d\sigma_n(t)\ \frac{1}{2\pi
i}\int_{-\infty}^{+\infty} \frac{dx}{x}\ \frac{1}{\det[AA^*
+(\varepsilon^2 - t)I -i\varepsilon \sqrt{t}(x+\frac{1}{x})I]}
\end{equation}
which is the identity claimed in Theorem \ref{thm2a}. It remains
to justify reversing the order of integrations with respect to $x$
and $\vec{v}$, $\vec{w}$. Firstly, we will show that the integral
on the right-hand side in (\ref{eq4:9}) is well-defined.

\smallskip

\begin{Prop}\label{sec4:prop2} For any $\varepsilon >0$ and $n\ge 2$ the integral in (\ref{eq4:9})
converges absolutely (and uniformly in $A$).
\end{Prop}

\emph{Proof.} Let $a_j^2$ be the eigenvalues of $AA^*$ so that
\[
\frac{1}{x}\times \frac{1}{\det [AA^* +(\varepsilon^2 - t)I
-i\varepsilon \sqrt{t}(x+\frac{1}{x})I]}= \frac{1}{x}\
\prod_{j=1}^n w(a_j,t)
\]
where
\[
w(a,t)= \frac{1}{a^2 +\varepsilon^2 - t -i\varepsilon
\sqrt{t}(x+\frac{1}{x})}.
\]
Since $|z| \ge | \im z|$,  we have
\[
\left|\frac{1}{x}\  w(a_1,t) \right|\le \frac{1}{\varepsilon
\sqrt{t}(1+x^2)}.
\]
Also, for all $0\le t \le \varepsilon^2/2$ we have
\[
|w(a_j,t)| \le \frac{1}{|\varepsilon^2 -t +a_j^2|}\le
\frac{2}{\varepsilon^2}
\]
and for all $\varepsilon^2/2 \le t \le1$ we have
\[
|w(a_j,t)| \le \frac{1}{|\varepsilon\sqrt{t} (x+\frac{1}{x})|} \le
\frac{1}{2\varepsilon\sqrt{t}} \le \frac{1}{\sqrt{2}
\varepsilon^2}.
\]
Therefore the absolute value of the integrand in (\ref{eq4:9}) is
majorated by the function
\[
\frac{1}{\varepsilon \sqrt{t}(1+x^2)}\
\left(\frac{2}{\varepsilon^2}\right)^{n-1},
\]
which is obviously integrable with respect to $d\sigma_n(t)\times
dx$. \hfill $\Box$

\medskip

We can now turn to justification of reversing the order of
integrations in $\vec{v},\vec{w}$ and $x$ in the integral
\begin{equation}\label{eq4:10}
{\cal I}=\int_0^1d\sigma_n(t) \int_{\C^n}\hspace{-1ex}\dd
\vec{v}\int_{\C^n}\hspace{-1ex}
 \dd \vec{w}\hspace{1ex} e^{ -\varepsilon
(\vec{v}^*\vec{v} +\vec{w}^*\vec{w}) +i(\vec{w}^* A^* \vec{v} +
\vec{v}^* A \vec{w})}\int_{-\infty}^{+\infty}
\frac{dx}{x}\hspace{1ex}
e^{i\sqrt{t}(x\vec{v}^*\vec{v}+\frac{\vec{w}^*\vec{w}}{x})}.
\end{equation}
%$-Q(\vec{v},\vec{w})$
The corresponding calculation is  routine but tedious. First we
restrict the $x$-integration to the finite interval $\delta \le
|x| \le 1/\delta$, $\delta >0$, reverse the order of integrations,
and then show that the corresponding tail integrals in are
negligible in the limit $\delta \to 0$.

Let
\[
{\cal I}_{\delta} = \int\limits_0^1d\sigma_n(t)
\int_{\C^n}\hspace{-1ex}\dd \vec{v}
\int\limits_{\C^n}\hspace{-1ex} \dd \vec{w} \int\limits_{\delta
\le |x| \le 1/\delta} \frac{dx}{x}\hspace{1ex} e^{ -\varepsilon
(\vec{v}^*\vec{v} +\vec{w}^*\vec{w}) +i(\vec{w}^* A^* \vec{v} +
\vec{v}^* A \vec{w})}\hspace{1ex}
e^{i\sqrt{t}(x\vec{v}^*\vec{v}+\frac{\vec{w}^*\vec{w}}{x})}
\]
The absolute value of the integrand is majorated by the integrable
function $\frac{1}{|x|} e^{-\varepsilon ( \vec{v}^*\vec{v}
+\vec{w}^*\vec{w})}$, and therefore we can reverse the order of
integrations  and then perform the integration in $\vec{v},
\vec{w}$. This yields
\[
{\cal I}_{\delta} = \int_0^1d\sigma_n(t)\int_{\delta \le |x| \le
1/\delta} \frac{dx}{x}\hspace{1ex} \frac{1}{\det [AA^*
+(\varepsilon^2 - t)I -i\varepsilon \sqrt{t}(x+\frac{1}{x})I]}.
\]
It follows from this, in view of By Proposition \ref{sec4:prop2},
that
\[
{\cal I}_{\delta}= \int_0^1d\sigma_n(t)\int_{-\infty}^{+\infty}
\frac{dx}{x}\hspace{1ex} \frac{1}{\det [AA^* +(\varepsilon^2 - t)I
-i\varepsilon \sqrt{t}(x+\frac{1}{x})I]} +o(1)
\]
in the limit $\delta \to 0$. It only remains to show that the tail
integrals
\[
{\cal I}_{\delta}^{\hspace{0.5ex}\prime} = \int_0^1d\sigma_n(t)
\int_{\C^n}\hspace{-1ex}\dd \vec{v} \int_{\C^n}\hspace{-1ex}
\dd\vec{w} \hspace{1ex} e^{ -\varepsilon (\vec{v}^*\vec{v}
+\vec{w}^*\vec{w}) +i(\vec{w}^* A^* \vec{v} + \vec{v}^* A
\vec{w})}\int_{|x|\ge 1/\delta } \frac{dx}{x} \hspace{1ex}
e^{i\sqrt{t}(x\vec{v}^*\vec{v}+\frac{\vec{w}^*\vec{w}}{x})}
\]
and
\[
{\cal I}_{\delta}^{\hspace{0.5ex}\prime\prime} =
\int_0^1d\sigma_n(t) \int_{\C^n}\hspace{-1ex}\dd \vec{v}
\int_{\C^n}\hspace{-1ex}   \dd \vec{w} \hspace{1ex} e^{ -\varepsilon
(\vec{v}^*\vec{v} +\vec{w}^*\vec{w}) +i(\vec{w}^* A^* \vec{v} +
\vec{v}^* A \vec{w})}\int_{|x|\le \delta } \frac{dx}{x}\hspace{1ex}
e^{i\sqrt{t}(x\vec{v}^*\vec{v}+\frac{\vec{w}^*\vec{w}}{x})}.
\]
vanish in the limit $\delta \to 0$.

For real $r,p$ and $q$ define
\begin{equation}\label{eq4:12}
g_L(r,p,q)=\int_{L}^{+\infty} \frac{dx}{x}\hspace{1ex}e^{ir
(px+\frac{q}{x})}= \int_{0}^{1/L} \frac{dx}{x}\hspace{1ex} e^{ir
(qx+\frac{p}{x})}.
\end{equation}
By integrating by parts,
\[
g_L(t;p,q) = -\frac{1}{iprL}e^{ir (pL+\frac{q}{L})} +
\frac{1}{ipr}\int_L^{+\infty} \frac{e^{ir (px+\frac{q}{x})}}{x^2}\
dx +\frac{p}{q} \int_L^{+\infty} \frac{e^{ir
(px+\frac{q}{x})}}{x^3}\ dx,
\]
and, therefore, for $L>0$ we have
\begin{equation}\label{eq4:11}
\left| g_L(r, p,q) \right| \le \frac{2}{|p|  |r| L}
+\frac{|q|}{2|p|L^2}.
\end{equation}
Obviously,
\[
\int_{|x|\ge 1/\delta } \frac{dx}{x}\hspace{1ex}
e^{i\sqrt{t}(x\vec{v}^*\vec{v}+\frac{\vec{w}^*\vec{w}}{x})} =
g_{\frac{1}{\delta}}(\sqrt{t},\vec{v}^*\vec{v},\vec{w}^*\vec{w})
 -
g_{\frac{1}{\delta}}(-\sqrt{t},\vec{v}^*\vec{v},\vec{w}^*\vec{w}),
\]
and, by (\ref{eq4:11}),
\[
\left| \int_{|x|\le \delta } \frac{dx}{x}\hspace{1ex}
e^{i\sqrt{t}(x\vec{v}^*\vec{v}+\frac{\vec{w}^*\vec{w}}{x})}
 \right| \le \frac{4\delta}{\vec{v}^*\vec{v}\sqrt{t}} +
\frac{\vec{w}^*\vec{w}\delta^2}{\vec{v}^*\vec{v}}.
\]
Therefore
\[
\left|{\cal I}_{\delta}^{\hspace{0.5ex}\prime} \right| \le
\int_0^1d\sigma_n(t) \int_{\C^n}\hspace{-1ex}\dd \vec{v}
\int_{\C^n}\hspace{-1ex}   \dd \vec{w} \hspace{1ex} e^{
-\varepsilon (\vec{v}^*\vec{v} +\vec{w}^*\vec{w})}
\left(\frac{4\delta}{\vec{v}^*\vec{v}\sqrt{t}} +
\frac{\vec{w}^*\vec{w}\delta^2}{\vec{v}^*\vec{v}} \right).
\]
As the function $\frac{1}{\vec{v}^*\vec{v}}$ is locally integrable
with respect to $\dd \vec{v} $ for $n\ge 2$, we conclude that
\begin{equation}\label{eq4:14}
{\cal I}_{\delta}^{\hspace{0.5ex}\prime}  = O(\delta) \hspace{3ex}
\hbox{when $\delta \to 0$}.
\end{equation}
Similarly
\[
\int_{|x|\le \delta } \frac{dx}{x}\hspace{1ex}
e^{i\sqrt{t}(x\vec{v}^*\vec{v}+\frac{\vec{w}^*\vec{w}}{x})} =
g_{\frac{1}{\delta}}(\sqrt{t},\vec{w}^*\vec{w},\vec{v}^*\vec{v})
 -
g_{\frac{1}{\delta}}(-\sqrt{t},\vec{w}^*\vec{w},\vec{v}^*\vec{v}),
\]
and repeating the above argument one obtains that
\[
{\cal I}_{\delta}^{\hspace{0.5ex}\prime\prime}  = O(\delta)
\hspace{3ex} \hbox{when $\delta \to 0$},
\]
so that both ${\cal I}_{\delta}^{\hspace{0.5ex}\prime} $ and
${\cal I}_{\delta}^{\hspace{0.5ex}\prime\prime}$ vanish in the
limit $\delta \to 0$. This completes our proof of the first part
of Theorem \ref{thm2a}.

It is worth mentioning another formula for the regularized average
of the inverse spectral determinant,
\[
R_{\varepsilon} (A,A^*) = (n-1)\int_{|z|^2 \le 1}
\frac{(1-|z|^2)^{n-2} \dd z}{\det [\varepsilon^2I + (1-\bar z)I +
(1-z)AA^*]},
\]
which is almost an immediate corollary of (\ref{eq3:7}) and the
representation
\begin{eqnarray*}
\frac{1}{\det[\varepsilon^2I_n + (I_n-AU)(I_n-AU)^*]} &\!\!=\!\!&
\frac{1}{\pi^n }\int_{\C^n} e^{-\vec{v}^* [\varepsilon^2I +
(I-AU)(I-AU)^*]\vec{v}} \dd \vec{v}\\[1ex]
 & \!\! = \!\! & \frac{1}{\pi^n }\! \int_{\C^n}\, e^{-[\vec{v}^*(1+\varepsilon^2) \vec{v}
 + \vec{v}^* AA^*\vec{v}]}\ e^{\tr (\vec{v}\vec{v}^* AU +
 U^* A^* \vec{v} \vec{v}^*)} \dd \vec{v}.
\end{eqnarray*}
This formula, however, does not seem to be easy to handle in the
limit $\varepsilon \to 0$.

We now turn to the integral in (\ref{eq4:9}) and evaluate it in
the limit $\varepsilon \to 0$ under the assumption that $AA^*$ has
no repeated eigenvalues. The limit in (\ref{ef}) follows
immediately from the asymptotic relation (\ref{fe}) which is the
end-product of our calculation.

The following identities, which can be obtained from the Lagrange
interpolation formula, see, e.g., \cite{PS}, will be useful for
our purposes.

\medskip

\begin{Prop} Suppose that $x_1, \ldots , x_n$ are pairwise distinct. Then
\begin{equation}\label{eq4:15}
\prod_{j=1}^n \frac{1}{x_j-t} = \sum_{j=1}^n \hspace{0.5ex}
\frac{1}{x_j-t} \hspace{0.5ex} \prod_{k\not=j} \frac{1}{x_k-x_j},
\end{equation}
and, for non-negative integer $r$,
\begin{equation}\label{eq4:16}
\sum_{j=1}^n \hspace{0.5ex} x^r_j \hspace{0.5ex} \prod_{k\not=j}
\frac{1}{x_k-x_j} =
  \begin{cases}
    0 & \text{if $r\le n-2$}, \\
    h_{r-n+1}(x_1, \ldots , x_n) & \text{if $r \ge n-1$},
  \end{cases}
\end{equation}
where the $h_r$, $r=0,1,2 ,\ldots ,$ are the complete symmetric
functions.
\end{Prop}

It follows from (\ref{eq4:15}) that
\begin{equation}\label{eq4:17}
\frac{1}{\det[AA^* +(\varepsilon^2 - t)I -i\varepsilon
\sqrt{t}(x+\frac{1}{x})I]} = \sum_{j=1}^n w(a_j,t,x)\
\prod_{k\not=j} \frac{1}{a_k^2-a_j^2},
\end{equation}
where $a_1, \ldots , a_n$ are the eigenvalues of $AA^*$ and
\[
w(a,t,x)= \frac{1}{a^2 +\varepsilon^2 - t -i\varepsilon
\sqrt{t}(x+\frac{1}{x})}.
\]
By the calculus of residues,
\begin{equation}\label{eq4:18}
\frac{1}{2\pi i }\int_{-\infty}^{+\infty} \frac{dx}{x}\ w(a,t, x)
=
\frac{1}{\sqrt{(a^2-t-\varepsilon^2)^2 + 4\varepsilon^2 a^2}},
\end{equation}
and putting (\ref{eq4:17}) and (\ref{eq4:18}) into (\ref{eq4:9})
we arrive at the following expression of $R_{\varepsilon} (A,A^*)$
in terms of the eigenvalues of $AA^*$:
\begin{equation}\label{eq4:19}
R_{\varepsilon} (A,A^*) = \sum_{j=1}^n \hspace{0.5ex}
F_{\varepsilon}(a_j) \hspace{0.5ex} \prod_{k\not=j}
\frac{1}{a_k^2-a_j^2},
\end{equation}
where
\begin{equation}\label{eq4:20}
F_{\varepsilon}(a)= \int_0^1
\frac{d\sigma_n(t)}{\sqrt{(a^2-t-\varepsilon^2)^2 + 4\varepsilon^2
a^2}}, \hspace{3ex} d\sigma_n(t)=(n-1)(1-t)^{n-2} dt.
\end{equation}
This formula is convenient for finding $R_{\varepsilon} (A,A^*)$
in the limit $\varepsilon \to 0$.

If $AA^* >I$, by letting $\epsilon \to 0$ in (\ref{eq4:19}) and
recalling (\ref{eq4:15}) we immediately obtain
\[
\lim_{\varepsilon \to 0}R_{\varepsilon} (A,A^*) = \int_0^1
\frac{d\sigma_n(t)}{\det (AA^*-tI)},
\]
thus reproducing the corresponding formula of Theorem \ref{thm1}
part (ii). If $AA^* <I$ or if $AA^*$ has eigenvalues on each side
of $a^2=1$, evaluation of the right-hand side in (\ref{eq4:19}) in
the limit $\varepsilon \to 0$  requires some work.

The integral in (\ref{eq4:20}) is standard. There are different
methods available to evaluate it. None seems to give an explicit
expression for all parameter values. However, we are only
interested in $\varepsilon \to 0$, and in this regime
\begin{equation}\label{eq4:21}
F_{\varepsilon}(a) =
(n-1)(1-a^2)^{n-2}\left[\gamma_{n-2}\sign(a^2-1) + L_0
(\varepsilon, a)\right] +q_{n-2}(a^2) +O(\varepsilon),
\end{equation}
where
\[
L_0 (\varepsilon, a) =
  \begin{cases}
    \ln \frac{1-a^2}{\varepsilon^2} & \text{if $a^2<1$}, \\
    \ln \frac{a^2}{a^2-1}  & \text{if $a^2>1$},\\
     \ln \frac{2}{\varepsilon} & \text{if
     $a^2=1$};
  \end{cases}
\]
$\gamma_{n-2}$ is the partial sum of the harmonic series,
\[
\gamma_{n-2} = \sum_{j=1}^{n-2} \frac{1}{j},
\]
$\sign$ is the sign function, $\sign (x)$ takes value 1 if $x>0$,
-1 if $x<0$ and 0 if $x=0$, and $q_{n-2} (a^2)$ is a polynomial of
degree $n-2$ in $a^2$ with coefficients which do not depend on
$\varepsilon$. Details of derivation of (\ref{eq4:21}) are given
in Appendix \ref{AppendixA}.

Let us now put (\ref{eq4:21}) into (\ref{eq4:19}). In view of
(\ref{eq4:16}) the polynomial $q_{n-2}$ gives no contribution.
After rearranging the remaining terms we obtain,
\begin{equation}\label{fe}
R_{\varepsilon} (A,A^*) = \alpha (AA^*) \ln
\frac{1}{\varepsilon^2} + \beta (AA^*) + O(\varepsilon),
\end{equation}
with the coefficients $\alpha$ and $\beta$ given by
\begin{eqnarray*}
 \alpha (AA^*) &=& (n-1) \sum_{j=1}^n \hspace{0.5ex} (1-a_j^2)^{n-2} \hspace{0.5ex}\theta
 (1-a_j^2) \hspace{0.5ex} \prod_{k\not=j}\hspace{0.5ex} \frac{1}{a_k^2-a_j^2}\\
\beta (AA^*) &=& (n-1) \sum_{j=1}^n \hspace{0.5ex} (1-a_j^2)^{n-2}
\hspace{0.5ex} \psi (a^2_j) \hspace{0.5ex}
 \prod_{k\not=j}\hspace{0.5ex} \frac{1}{a_k^2-a_j^2}
\end{eqnarray*}
where $\theta$ is Heaviside's step function,
\[
\theta (x) =
  \begin{cases}
    1 & \text{if $x>1$}, \\
    \frac{1}{2} & \text{if $x=1$}\\
    0 & \text{if $x<1$}
  \end{cases}
\]
and
\[
\psi(a^2)= \begin{cases}
    \gamma_{n-2} +\ln a^2 - \ln (a^2-1)  & \text{if $a^2>1$}, \\
    -\gamma_{n-2} + \ln (1-a^2) & \text{if $a^2<1$}\\
    \ln 2 & \text{if $a^2=1$}.
  \end{cases}
\]

As one would expect, the coefficient $\alpha$ vanishes if $AA^*<I$
or $AA^*>I$. This follows from identity (\ref{eq4:16}). If
$AA^*>I$ then the constant $\gamma_{n-2}$ gives no contribution,
again by (\ref{eq4:16}) and
\[
\lim_{\varepsilon \to 0}R_{\varepsilon} (A,A^*) = (n-1)
\sum_{j=1}^n \hspace{0.5ex} (1-a_j^2)^{n-2} \hspace{0.5ex} \ln
\frac{a_j^2}{a_j^2-1} \hspace{0.5ex}
 \prod_{k\not=j}\hspace{0.5ex} \frac{1}{a_k^2-a_j^2} = \int_0^1
 \frac{d\sigma_n(t)}{\det(AA^*-tI)}.
\]
Similarly,  if $AA^*<I$ then
\[
\lim_{\varepsilon \to 0}R_{\varepsilon}  (A,A^*) = (n-1)
\sum_{j=1}^n \hspace{0.5ex} (1-a_j^2)^{n-2} \hspace{0.5ex} \ln (1-
a^2_j) \hspace{0.5ex}
 \prod_{k\not=j}\hspace{0.5ex} \frac{1}{a_k^2-a_j^2} = \int_0^1
 \frac{d\sigma_n(t)}{\det(I-tAA^*)}.
\]
Thus, (\ref{eq4:9}) indeed reproduces formulas of Theorem
\ref{thm1} part (ii).

\section{Rank-one perturbations of CUE and GUE} \label{sec6}

In this section we express the mean eigenvalue density for the
random matrix ensembles (\ref{con}) and (\ref{res}) in terms of
the spectral determinants. Our calculation  is inspired by similar
calculations in \cite{E,EKS} and makes use of a process known as
eigenvalue deflation which was introduced in the context of random
matrices in \cite{Trotter}.

We need to recall a few facts about elementary unitary Hermitian
matrices \cite{Wilkinson}. Let $\vec{v}$ be a column-vector in
$\C^n$. The matrix
\[
R_{\vec{v}} = I_n-2\vec{v} \vec{v}^*/|\vec{v}|^2, \hspace{3ex}
|\vec{v}|^2= \vec{v}^*\vec{v},
\]
where $I_n$ is identity matrix, defines a linear transformation
which is reflection across the hyperplane through the origin with
normal $\vec{v}/|\vec{v}|$. It is straightforward to verify that
$R_{\vec{v}}$ is unitary and Hermitian,
\[
R_{\vec{v}}=R_{\vec{v}}^* \hspace{1ex} \hbox{and} \hspace{1ex}
R_{\vec{v}}R_{\vec{v}}^*=R_{\vec{v}}^2=I_n.
\]
In the context of numerical linear algebra the matrices
$R_{\vec{v}}$ are known as Householder reflections. Any matrix can
be brought to triangular form by a succession of Householder
reflections. We only need the first step of this process which we
now describe.

Let $W_n$ be an $n\times n$ matrix and $z$ and
$\vec{x}=(x_1,\ldots, x_n)^T$ be an eigenvalue and eigenvector of
$W_n$, so that
\[
W_n \vec{x}=z\vec{x}.
\]
Without loss of generality we may assume that $x_1\ge 0$ and
$|\vec{x}|^2=\vec{x}^*\vec{x} = 1$. Let $\vec{e_1}=(1, 0, \ldots ,
0)^T$ and
\begin{equation}\label{eq5:1}
\vec{v}= \frac{\vec{x} + \vec{e_1}}{|\vec{x} +
\vec{e_1}|}=\frac{\vec{x} + \vec{e_1}}{\sqrt{2(1+x_1)}}.
\end{equation}
Since the vector $\vec{v}$ bisects the angle of $\vec{x}$ and
$\vec{e_1}$, we have $R_{\vec{v}} \vec{x}=-\vec{e_1}$ and
$R_{\vec{v}} \vec{e_1}= - \vec{x}$. Therefore $R_{\vec{v}} W_n
R_{\vec{v}}\vec{e_1}=z\vec{e_1}$ and (recall that
$R_{\vec{v}}^2=I_n$)
\begin{equation}\label{eq5:2}
 W_n = R_{\vec{v}}
\begin{pmatrix}
  z & \vec{w}^* \\
  \vec{0} & W_{n-1}
\end{pmatrix} R_{\vec{v}},
\end{equation}
for some $W_{n-1}$ and $\vec{w}$. Note that $W_{n-1}$ is
$(n-1)\times (n-1)$ and $\vec{w}^*$ is $1\times (n-1)$. Obviously,
applying this procedure again (to the matrix $W_{n-1}$) and again,
one can reduce $W_n$ to triangular form by means of unitary
transformations. Such factorization is known as Schur
decomposition.

It is convenient to write $\vec{v}=(v_1, \vec{q})^T$, where
$\vec{q}=(v_2, \ldots, v_n)^T$. Since $\vec{v}$ is a unit vector,
$v_1^2+|\vec{q}|^2=1$. Note that the first equation in
$(\ref{eq5:1})$ reads $v_1=\sqrt{(1+x_1)/2}$. Since $0\le x_1 \le
1$, we must have $1/2 \le v_1\le 1$. Therefore,
\begin{equation}\label{eq5:sph}
v_1= \sqrt{1-|\vec{q}|^2} \hspace{2ex}\hbox{and}\hspace{2ex}
\frac{1}{2}\le |\vec{q}|^2\le 1.
\end{equation}
In terms of $\vec{q}$ the matrix $R_{\vec{v}}$ is given by
\begin{equation}\label{eq5:16}
R_{\vec{v}} = \begin{pmatrix}
  1-2v_1^2 & -2v_1\vec{q}^* \\
  -2v_1\vec{q} & I_{n-1} - 2\vec{q}\vec{q}^*
\end{pmatrix} = \begin{pmatrix}
  2|\vec{q}|^2-1& -2\sqrt{1-|\vec{q}|^2}\vec{q}^* \\
  -2\sqrt{1-|\vec{q}|^2}\vec{q} & I_{n-1} - 2\vec{q}\vec{q}^*
\end{pmatrix} .
\end{equation}

The incomplete Schur decomposition (\ref{eq5:2}) gives rise to a
new coordinate system in the space of complex matrices, the new
(complex) coordinates being $z$, $\vec{w}$, $\vec{q}$ and the
matrix entries of $W_{n-1}$. There are no restrictions on the
range of variation of $z$, $\vec{w}$ and $W_{n-1}$, and, in view
of (\ref{eq5:sph}), the vector $\vec{q}$ is restricted to the
spherical segment $\frac{1}{2}\le |\vec{q}|^2 \le 1$. The Jacobian
of the transformation from $(W_{n,jk})$ to this new system of
coordinates\footnote{For Jacobian computations it is convenient to
consider $z$ and $\overline{z}$ as functionally independent
variables, so that $d^2z\equiv d\re z d\im z
=dzd\overline{z}/2$.},
\begin{eqnarray*}
\lefteqn{\prod_{j,k=1}^n \frac{d(W_n)_{jk}
d\overline{(W_n)}_{jk}}{2} =} \\ & &  J(z, \vec{w}, \vec{q},
W_{n-1})\ \frac{dz d\overline{z}}{2} \ \prod_{j=1}^{n-1}
\frac{dw_{j} d\overline{w_{j}}}{2}\ \prod_{j=1}^{n-1} \frac{dq_{j}
d\overline{q_{j}}}{2}\ \prod_{j,k=1}^{n-1} \frac{d(W_{n-1})_{jk}
d\overline{(W_{n-1})_{jk}}}{2},
\end{eqnarray*}
is given by (cf. Lemma 3.2 in \cite{EKS})
\begin{equation}\label{eq5:11}
J(z, \vec{w}, \vec{q}, W_{n-1}) = 2^{2n-2}|\det(zI_{n-1}
-W_{n-1})|^2(1-|\vec{q}|^2)^{n-2}(2|\vec{q}|^2-1).
\end{equation}
We derive (\ref{eq5:11}) in Appendix \ref{AppendixB}.

Suppose that we have a probability distribution
\begin{equation}\label{eq5:00}
dP(W_n)=p(W_n) \prod_{i,j=1}^n
\frac{d(W_{n})_{ij}d\overline{(W_{n})_{ij}}}{2}
\end{equation}
on the space of complex $n\times n$ matrices. Then, following the
argument of \cite{EKS}, see their Lemma 3.1, the mean eigenvalue
density $\rho_n(x,y)$, $z=x+iy$, of $W_n$ is given by
\begin{equation}\label{eq5:15n}
\rho_n(x,y) = \int\limits_{\C^{2(n-1)^2}} \hspace*{-2ex}
d^2W_{n-1}\int\limits_{\C^{n-1}}\hspace*{-0.5ex} d^2\vec{w}
\hspace*{-2ex}\int\limits_{\frac{1}{2}\le |\vec{q}|^2\le
1}\hspace*{-2ex} d^2\vec{q} \hspace*{1ex} J(z, \vec{w}, \vec{q},
W_{n-1}) p\left(R_{\vec{v}}
\begin{pmatrix}
  z & \vec{w}^* \\
  \vec{0} & W_{n-1}
\end{pmatrix} R_{\vec{v}}  \right),
\end{equation}
where
\[
d^2W_{n-1} = \prod_{j,k=1}^{n-1} \frac{d(W_{n-1})_{jk}
d\overline{(W_{n-1})_{jk}}}{2}, \hspace*{2ex}
d^2\vec{q}=\prod_{j=1}^{n-1} \frac{dq_{j} d\overline{q_{j}}}{2},
\hspace*{2ex} d^2\vec{w} = \prod_{j=1}^{n-1} \frac{dw_{j}
d\overline{w_{j}}}{2}.
\]
%The Jacobian $J(z, \vec{w}, \vec{q}, A_{n-1})$ vanishes when one
%of the eigenvalues of $A_0$ coincides with $z$ or when $x_1=0$.
%With our choice of matrix distributions, these events will have
%probability zero, and we can neglect them.

Since we integrate in the $\vec{q}$-space over the spherical segment
$1/2 \le |\vec{q}|^2\le 1$, it is convenient to introduce spherical
coordinates,
\[
\vec{q}=\sqrt{t}\vec{\sigma}, \hspace{2ex} t=|\vec{q}|^2,
\hspace{2ex} \vec{\sigma} = \vec{q}/|\vec{q}|.
\]
The element of volume in the $\vec{q}$-space is then
\[
d^2\vec{q}=\frac{1}{2}\, t^{n-2}\, dt\, dS(\vec{\sigma}),
\]
where $dS(\vec{\sigma})$ is the element of area of the sphere
$|\vec{\sigma}|^2=1$. The range of variation of $t$ is $1/2 \le t
\le 1$. Next, on making the substitution
\[
r=(2t-1)^2, \hspace{2ex} (2t-1)dt=\frac{1}{4}dr, \hspace{2ex}
(1-t)t=\frac{1}{4}(1-r), \hspace{2ex} t=\frac{1+\sqrt{r}}{2},
\]
the expression for the Jacobian becomes simpler,
\begin{eqnarray*}
J(z,\vec{w}, \vec{q}, W_{n-1})d^2\vec{q}&=&J(z,\vec{w}, \sqrt{t}
\vec{\sigma}, W_{n-1}) d^2\vec{q}\\ &=& 2^{2n-3}|\det(zI_{n-1}
-W_{n-1})|^2[(1-t)t]^{n-2}(2t-1) dt dS(\vec{\sigma})\\ & =&
\frac{1}{2} |\det(zI_{n-1} -W_{n-1})|^2(1-r)^{n-2}dr
dS(\vec{\sigma}).
\end{eqnarray*}
Substituting this into (\ref{eq5:15n}), we arrive at the desired
formula for the mean density of eigenvalues in the ensemble with
matrix distribution (\ref{eq5:00}):
\begin{equation}\label{eq5:15}
\rho_n(x,y) \!=\!
\frac{1}{2}\hspace{-2ex}\int\limits_{\C^{2(n-1)^2}} \hspace*{-3ex}
d^2W_{n-1}\hspace*{-2ex} \int\limits_{\C^{n-1}}\hspace*{-1ex}
d^2\vec{w} \hspace{-2ex}
\int\limits_{|\vec{\sigma}|^2=1}\hspace{-2ex}
dS(\vec{\sigma})\hspace{-1ex}\int\limits_{0}^1 \!(1\!-\!r)^{n-2}
dr |\det(zI_{n-1}\!-\!W_{n-1})|^2 p\left(\!R
\begin{pmatrix}
  z & \vec{w}^* \\
  \vec{0} & W_{n-1}
\end{pmatrix}\!\! R\! \right).
\end{equation}
Here
\begin{equation}
R=\left(
    \begin{array}{cc}
      \sqrt{r} & \sqrt{1-r}\ \vec{\sigma}^* \\[2ex]
      -\sqrt{1-r}\ \vec{\sigma} & I_{n-1}-(1+\sqrt{r})\vec{\sigma}\vec{\sigma}^* \\
    \end{array}
  \right).
\end{equation}

We shall now apply this result to express the mean density of
eigenvalues in terms of the absolute square modulus of
characteristic polynomials for two ensembles of random matrices.

\medskip

\noindent {\bf Rank-one deviations from unitarity.} \hspace{1ex}
Let $U_n$ be an $n\times n$ unitary matrix and
\begin{equation}\label{eq5:14}
G_n= \begin{pmatrix}
  \sqrt{1-\gamma} & \vec{0} \\
  \vec{0} & I_{n-1}
\end{pmatrix}, \hspace{3ex} 0\le  \gamma  \le 1.
\end{equation}
The Haar measure on $U(n)$ induces a measure on the manifold
$W_n^*W_n=G_n^2$ in the space of complex $n\times n$ matrices via
the correspondence $W_n=U_nG_n$. The corresponding matrix
distribution is uniform and can be conveniently described via
matrix delta-function
\begin{equation}\label{eq5:12}
dP(W_n)= c_n\delta (W_n^*W_n - G_n^2) \prod_{i,j=1}^n
\frac{d(W_n)_{ij}d\overline{(W_n)_{ij}}}{2}.
\end{equation}
For Hermitian $H$, we define $\delta (H)$ as
\begin{equation}\label{delta}
\delta (H)=
\prod_{j} \delta (H_{jj}) \prod_{j<l} \delta (H_{jk}) \delta
(\overline{H_{jk}}).
\end{equation}
The normalization constant $c_n$ can be easily computed by
changing to the matrix `polar' coordinates,
\begin{equation}\label{eq5:13}
c_n=\frac{2^n}{\mathrm{Vol} (U(n))}= \frac{1!2! \cdots
(n-1)!}{\pi^{n(n+1)/2}}.
\end{equation}
Note that the eigenvalues of $G_nU_n$ and $U_nG_n$ coincide, and,
therefore, for the purpose of calculating the eigenvalue
statistics the ensembles $G_nU_n$ and $U_nG_n$ are equivalent. In
view of (\ref{eq5:2}) it is more convenient to work with matrices
$U_nG_n$.

Changing the coordinate system to $z$, $\vec{w}$, $r$,
$\vec{\sigma}$, $W_{n-1}$,
\begin{eqnarray} \label{eq5:17}
dP(W_n)&=&\frac{c_n}{2}\  \delta\left(\begin{pmatrix}
  \overline{z} & \vec{0} \\
  \vec{w} & W_{n-1}^*
\end{pmatrix} \begin{pmatrix}
  z & \vec{w}^* \\[1ex]
  \vec{0} & W_{n-1}
\end{pmatrix} - RG_n^2R\right) \times \\ \nonumber & &
 |\det(zI_{n-1}\!-\!W_{n-1})|^2 d^2z\ (1-r)^{n-2} dr\
dS(\vec{\sigma})\ d^2\vec{w}\ d^2W_{n-1},
\end{eqnarray}
where we have used the unitary invariance of the matrix
delta-function.

The matrix inside the delta-function in (\ref{eq5:17}) is
\[
\begin{pmatrix}
  |z|^2 - 1+\gamma r & \overline{z}\vec{w}^* - \gamma
\sqrt{(1-r)r}\ \vec{\sigma}^* \\[2ex]
  z\vec{w} - \gamma
\sqrt{(1-r)r}\ \vec{\sigma} & W_{n-1}^*W_{n-1}-I_{n-1} +
\vec{w}\vec{w}^*
 +\gamma (1-r)\vec{\sigma}\vec{\sigma}^*
\end{pmatrix}
\]
and the delta-function factorizes into the product of the three
delta-functions, correspondingly. On substituting this into
(\ref{eq5:15}) we obtain
\begin{equation}\label{eq5:20}
\rho_n(x,y) = \frac{c_n}{2} \int\limits_{\C^{2(n-1)^2}}
\hspace*{-2ex} |\det(zI_{n-1} -W_{n-1})|^2 f(W_{n-1}) \ d^2W_{n-1}
\end{equation}
where
\begin{eqnarray*}
\!\!f(W_{n-1})\!\!&\!\! = \!\!& \!\!
\int\limits_{|\vec{\sigma}|^2=1}\hspace{-2ex} dS(\vec{\sigma})
 \int\limits_0^1 (1-r)^{n-2} dr\
\delta \left(|z|^2 \!- \!1\!+\!\gamma r \right) \times \\
\!\!&\!\! &\!\! \int\limits_{\C^{n-1}} d^2\vec{w}\ \delta
\left(W_{n-1}^*W_{n-1}\! -\!I_{n-1}\!+\!\vec{w}\vec{w}^* \!+\!
\gamma (1\!-\!r)\vec{\sigma}\vec{\sigma}^*\right) \delta
\left(z\vec{w}\! -\! \gamma \sqrt{(1\!-\!r)r}\
\vec{\sigma}\right).
\end{eqnarray*}
Note that
\[
\delta  \left(z\vec{w} - \gamma \sqrt{(1-r)r}\ \vec{\sigma}\right) =
\frac{1}{|z|^{2(n-1)}}\  \delta \left(\vec{w} - \frac{\gamma
\sqrt{(1-r)r}\ \vec{\sigma}}{z}\right).
\]
Therefore the integral over $\vec{w}$ yields
\[
\frac{1}{|z|^{2(n-1)}}\ \delta \left(W_{n-1}^*W_{n-1}  -I_{n-1} +
\frac{\gamma (1-r)(\gamma r + |z|^2)}{|z|^2}\
\vec{\sigma}\vec{\sigma}^* \right)
\]
and
\begin{eqnarray*}
f(W_{n-1}) &= &\frac{1}{|z|^{2(n-1)}}
\int\limits_{|\vec{\sigma}|^2=1}\hspace{-2ex} dS(\vec{\sigma})
 \int\limits_0^1 (1-r)^{n-2} dr\ \delta
\left(|z|^2 - 1+\gamma r\right) \times  \\ & &
\hspace*{20ex}\delta \left(W_{n-1}^*W_{n-1} -I_{n-1} +
\frac{\gamma (1-r)(\gamma r + |z|^2)}{|z|^2}\
\vec{\sigma}\vec{\sigma}^* \right).
\end{eqnarray*}
It is apparent that  if $|z|^2 < 1-\gamma$ or $|z|^2>1$ then the
integral over $r$ vanishes and $f(W_{n-1})= 0$. Therefore
$\rho_n(x,y)=0$ for such values of $z$. If $1-\gamma < |z|^2 < 1$
then the integral over $r$ produces a non-trivial contribution and
\[
f(W_{n-1})= \frac{1}{\gamma |z|^{2(n-1)}}\hspace{-1ex}
\int\limits_{|\vec{\sigma}|^2=1} \hspace{-1ex} dS(\vec{\sigma})\!
\left(\!1\!-\!\frac{1\!-\!|z|^2}{\gamma}\!\right)^{n-2}\!\! \delta
\left(\!W_{n-1}^*W_{n-1} \!-\!I_{n-1}\! +\!
\frac{\gamma\!-\!1\!+\!|z|^2}{|z|^2}\ \vec{\sigma}\vec{\sigma}^*\!
\right).
\]
Introducing
\[
\tilde \gamma = \frac{\gamma-1+|z|^2}{|z|^2},
\]
we can rewrite the above expression in a shorter form,
\[
f(W_{n-1})= \frac{1}{\gamma
|z|^2}\left(\frac{\tilde\gamma}{\gamma}\right)^{n-2}\hspace{-1ex}
\int\limits_{|\vec{\sigma}|^2=1} \hspace{-1ex} dS(\vec{\sigma})\
\delta \left(\!W_{n-1}^*W_{n-1} -I_{n-1} + \tilde \gamma
\vec{\sigma}\vec{\sigma}^*\! \right).
\]
On substituting this into (\ref{eq5:20}), we arrive at
\begin{eqnarray*}
\lefteqn{\rho_n(x,y) = \frac{c_n}{2\gamma |z|^2}\left(\frac{\tilde
\gamma }{\gamma}\right)^{n-2}
\hspace*{-2ex}\int\limits_{\C^{2(n-1)^2}}\hspace*{-1ex} d^2W_{n-1}
|\det(zI_{n-1}-W_{n-1})|^2\times} \\ & & \hspace*{25ex}
\int\limits_{|\vec{\sigma}|^2=1}\hspace*{-1ex} dS(\vec{\sigma})\
\delta \left(W_{n-1}^*W_{n-1} -I_{n-1} + \tilde \gamma
\vec{\sigma}\vec{\sigma}^* \right).
\end{eqnarray*}
Since the matrix $\vec{\sigma}\vec{\sigma}^*$ is unitary equivalent
to the matrix
\[
\begin{pmatrix}
  1 & \vec{0} \\
  \vec{0} & 0_{n-1}
\end{pmatrix},
\]
the integration over $\vec{\sigma}$ can be easily performed yielding
\[
\rho_n(x,y)= \frac{c_n\mathrm{Vol}(S^{2n-3})}{2\gamma
|z|^2}\left(\frac{\tilde \gamma }{\gamma}\right)^{n-2}
\hspace*{-3ex}\int\limits_{\C^{2(n-1)^2}}\hspace*{-1ex}
|\det(zI_{n-1}-W_{n-1})|^2 \delta \left(W_{n-1}^*W_{n-1} - {\tilde
G_{n-1}}^2 \right)\! d^2W_{n-1},
\]
where $\tilde G_{n-1}$ is the $(n-1)\times (n-1)$ matrix (cf.
(\ref{eq5:14}))
\[
\tilde G_{n-1}= \left(
            \begin{array}{cc}
              \sqrt{1-\tilde \gamma} & \vec{0} \\
              \vec{0} & I_{n-2} \\
            \end{array}
          \right)
\]
and $\mathrm{Vol}(S^{2n-3})$ is the area of the unit sphere in
$\R^{2(n-1)}$,
\begin{equation}\label{area}
\mathrm{Vol}(S^{2n-3})=\frac{2\pi^{n-1}}{(n-2)!}.
\end{equation}
It follows from (\ref{eq5:13}) that $c_n/c_{n-1}= (n-1)!/\pi^n$.
Hence
\[
\frac{c_n\mathrm{Vol}(S^{2n-3})}{2} = \frac{n-1}{\pi}c_{n-1},
\]
and finally
\begin{eqnarray*}
\rho_n(x,y) &=& \frac{n-1}{\pi \gamma |z|^2}\left(\frac{\tilde
\gamma }{\gamma}\right)^{n-2}\hspace*{-2ex} c_{n-1}
\hspace*{-2ex}\int\limits_{\C^{2(n-1)^2}}\hspace*{-1ex}
|\det(zI_{n-1}-W_{n-1})|^2 \delta \left(W_{n-1}^*W_{n-1} - {\tilde
G_{n-1}}^2 \right)\ d^2W_{n-1} \\ &=& \frac{n-1}{\pi \gamma
|z|^2}\left(\frac{\tilde \gamma
}{\gamma}\right)^{n-2}\hspace*{-1ex} \int_{U(n-1)}
\left|\det\left(zI_{n-1}-U_{n-1}\tilde G_{n-1}\right)\right|^2
dU_{n-1}
\end{eqnarray*}
as claimed in (\ref{int:eq2}).

\medskip

\noindent {\bf Rank-one deviations from Hermiticity.} \hspace{1ex}
Let $H_n$ be a GUE$_n$ matrix, i.e. random Hermitian matrix of
size $n\times n$ with probability distribution
\[
c_{\beta,n} e^{-\frac{\beta}{2} \tr H_n^2}\  \prod_{j=1}^n
d(H_n)_{jj} \prod_{1\le j<k\le n}
\frac{d(H_n)_{jj}d\overline{(H_n)_{jj}} }{2}, \hspace{3ex} \beta
>0,
\]
and $\Gamma_n$ be the $n\times n$ matrix
\[
\Gamma_n = \gamma \begin{pmatrix}
   1 & \vec{0} \\
   \vec{0} & 0_{n-1}
\end{pmatrix}, \hspace{3ex} \gamma >0
\]
Consider the random matrices
\[
W_n=H_n+i\Gamma_n.
\]
Obviously,
\[
\re W_n := \frac{W_n+W_n^*}{2}=H_n, \hspace{3ex} \hbox{and}
\hspace{3ex} \im W_n := \frac{W_n-W_n^*}{2i}=\Gamma_n.
\]
The matrices $W_n$ are complex and their probability distribution
is given by
\begin{equation}\label{eq5:12h}
dP(W_n)= c_{\beta,n}e^{-\frac{\beta}{2} \tr (\re W_n)^2} \delta
(\im W_n - \Gamma_n) \prod_{i,j=1}^n
\frac{d(W_n)_{ij}d\overline{(W_n)_{ij}}}{2},
\end{equation}
where $c_{\beta,n}$ is the normalization constant,
\begin{equation}\label{eq5:13h}
c_{\beta,n}=
\left(\frac{1}{2}\right)^{n/2}\left(\frac{\beta}{\pi}\right)^{n^2/2}
\end{equation}
and $\delta$ is the matrix delta-function (\ref{delta}).

Changing the coordinate system to $z$, $\vec{w}$, $r$,
$\vec{\sigma}$ and $W_{n-1}$,
\begin{eqnarray} \label{eq5:17h}
dP(W_n)\!\!&=&\!\!\frac{1}{2}c_{\beta,n}e^{-\frac{\beta}{2}\tr
(\re W_{n-1})^2 -\frac{\beta}{4} |\vec{w}|^2 -\frac{\beta}{2}(\re
z)^2 } \delta\left(\!\!\begin{pmatrix}
  \im z & \frac{\vec{w}^*}{2i} \\[1ex]
  \!\!-\frac{\vec{w}}{2i} & \im W_{n-1}
\end{pmatrix} \!-\! R\Gamma_n R\!\right)\!\times  \\ \nonumber \!\!& &\!\!
|\det(zI_{n-1}\!-\!W_{n-1})|^2 d^2z\ (1-r)^{n-2} dr\
dS(\vec{\sigma})\ d^2\vec{w}\ d^2W_{n-1},
\end{eqnarray}
where we have used the unitary invariance of the matrix
delta-function.

The matrix inside the delta-function in (\ref{eq5:17}) is
\[
\begin{pmatrix}
  \im z - \gamma r & \frac{\vec{w}^*}{2i} + \frac{\gamma}{2}
\sqrt{(1-r)r}\ \vec{\sigma}^* \\[1ex]
  -\frac{\vec{w}^*}{2i} + \frac{\gamma}{2}
\sqrt{(1-r)r}\ \vec{\sigma}^* & \im W_{n-1} - \gamma
(1-r)\vec{\sigma}\vec{\sigma}^*
\end{pmatrix}
\]
and the delta-function factorizes into the product of the three
delta-functions correspondingly. On substituting this into
(\ref{eq5:15}) we obtain
\begin{equation}\label{eq5:20h}
\rho_n(x,y) = \frac{1}{2} c_{\beta,n} e^{-\frac{\beta
x^2}{2}}\hspace*{-2ex} \int\limits_{\C^{2(n-1)^2}} \hspace*{-2ex}
|\det(zI_{n-1} -W_{n-1})|^2 f(W_{n-1}) e^{-\frac{\beta}{2} \tr
(\re W_{n-1})^2}\ d^2W_{n-1}
\end{equation}
where
\begin{eqnarray*}
f(W_{n-1})\!\!& = &\!\!\hspace*{-2ex} \int\limits_{
|\vec{\sigma}|^2=1 } \hspace{-2ex} dS(\vec{\sigma})
\int\limits_0^1 (1-r)^{n-2} dr \hspace*{1ex} \delta \left(y -
\gamma r \right)  \delta \left( \im W_{n-1} - \gamma
(1-r)\vec{\sigma}\vec{\sigma}^* \right) \times \\ \!\!& &\!\!
\hspace*{25ex} \int\limits_{\C^{n-1}}\hspace*{-1ex} d^2\vec{w}\
e^{-\frac{\beta |\vec{w}|^2}{4}} \delta \left(\frac{\vec{w}^*}{2i}
+ \frac{\gamma}{2} \sqrt{(1\!-\!r)r}\ \vec{\sigma}^*\right).
\end{eqnarray*}
The integral over $\vec{w}$ yields $\frac{1}{4}\ e^{-\beta
\gamma^2(1-r)r}$, and we arrive at
\[
f(W_{n-1}) = \frac{1}{4}\hspace{-1ex}
\int\limits_{|\vec{\sigma}|^2=1}\hspace{-2ex} dS(\vec{\sigma})
\hspace{-1ex} \int\limits_{0}^{1}\hspace{-1ex} dr  \ (1-r)^{n-2}
e^{-\beta \gamma^2 (1-r)r} \delta \left(y \!-\! \gamma r \right)
\delta \left(\im W_{n-1} \!- \!\gamma (1-r)
\vec{\sigma}\vec{\sigma}^* \! \right).
\]
It is apparent that if $y<0$ or $y>\gamma$ then the integral over
$r$ vanishes. Therefore $\rho_n(x,y)=0$ if $y<0$ or $y>\gamma$. If
$0<y<\gamma$, then the integration over $r$ produces the factor
$\frac{1}{\gamma}$ and the constraint $r=\frac{y}{\gamma}$, so
that
\[
f(W_{n-1}) = \frac{(\gamma - y)^{n-2} e^{-\beta
(\gamma-y)y}}{4\gamma^{n-1}}\hspace{-1ex}
\int\limits_{|\vec{\sigma}|^2=1}\hspace{-1ex} dS(\vec{\sigma})\
\delta \left(\im W_{n-1} - (\gamma -y) \vec{\sigma}\vec{\sigma}^*
\right).
\]
On substituting the obtained expression for $f(W_{n-1})$ into
(\ref{eq5:20h}), we arrive at
\begin{eqnarray*}
\rho_n(x,y)\!\!&\!\!=\!\!&\!\!
\frac{c_{\beta,n}(\gamma-y)^{n-2}e^{-\frac{\beta x^2}{2}-\beta
(\gamma-y)y}}{8\gamma^{n-1}}
\hspace*{-1ex}\int\limits_{\C^{2(n-1)^2}}\hspace*{-1ex} d^2W_{n-1}
|\det(zI_{n-1}\!-\!W_{n-1})|^2 e^{-\frac{\beta}{2} \tr (\re
(W_{n-1})^2 }\! \times \\ \!\!&\!\! &\!\!
\int\limits_{|\vec{\sigma}|^2=1}\hspace*{-1ex} dS(\vec{\sigma})\
\delta \left(\im W_{n-1} \!-\! (\gamma -y)
\vec{\sigma}\vec{\sigma}^* \right).
\end{eqnarray*}
The integral over $\vec{\sigma}$ yields
\[
\mathrm{Vol} (S^{2n-3})\ \delta \left(\im W_{n-1} - \tilde
\Gamma_{n-1} \right),
\]
where $\tilde \Gamma_{n-1}$ is the $(n-1)\times (n-1)$ matrix
\[
\tilde \Gamma_{n-1}= (\gamma - y)
\begin{pmatrix}
  1 & \vec{0} \\
  \vec{0} & 0_{n-2}
\end{pmatrix}
\]
and  $\mathrm{Vol} (S^{2n-3})$ is the area of the unit sphere in
$\R^{2(n-1)}$ (\ref{area}). We have that
\[
\frac{c_{\beta,n}}{c_{\beta,n-1}}=\left(\frac{\beta}{\pi}\right)^n
\left(\frac{\pi}{2\beta}\right)^{1/2}
\]
and it now follows that
\begin{eqnarray*}
\rho_n(x,y)\!\!&\!\!=\!\!&\!\!\frac{1}{4\sqrt{2\pi\beta}}\frac{\beta^n}{
(n-2)!} \frac{(\gamma-y)^{n-2}e^{-\frac{\beta x^2}{2}-\beta
(\gamma-y)y}}{\gamma^{n-1}}\times \\[1ex] \!\!&\!\! &
\!\!c_{\beta,n-1}
\hspace*{-1ex}\int\limits_{\C^{2(n-1)^2}}\hspace*{-1ex} d^2W_{n-1}
|\det(zI_{n-1}-W_{n-1})|^2 e^{-\frac{\beta}{2} \tr (\re
(W_{n-1})^2 } \delta \left(\im W_{n-1} - \tilde \Gamma_{n-1}
\right),
\end{eqnarray*}
as claimed in (\ref{int:eq2h}).

\appendix

\section{Appendix} \label{AppendixA}

In this appendix we evaluate the integral
\[
{\cal I}_k (\varepsilon^2, a^2) = \int_{0}^1 \frac{(1-t)^k\ dt
}{\sqrt{(t-a^2+\varepsilon^2)^2 + 4 \varepsilon^2 a^2 }}
\]
in the limit $\varepsilon \to 0$.

We shall use the following fact from Calculus. If $P(t)$ is a
polynomial of degree $k$ then (integrate by parts)
\begin{equation}\label{eqa:1}
\int \frac{P(t)\ dt }{\sqrt{t^2+pt + q}} = Q(t) \sqrt{t^2+pt + q}
+ \lambda \int \frac{dt}{\sqrt{t^2+pt + q}}
\end{equation}
where $Q$ is a polynomial of degree $k-1$ and $\lambda$ is a
constant. For $Q$ and $\lambda$ one has the equation
(differentiate (\ref{eqa:1}))
\[
P(t)=Q^{\prime}(t) (t^2+pt+q) + \frac{1}{2}Q(t)
(t^2+pt+q)^{\prime} +\lambda.
\]
It follows from this that
\begin{eqnarray*}
{\cal I}_k (\varepsilon^2, a^2) &=&
\left[Q_{\varepsilon,a}(t)\sqrt{(t-a^2
+\varepsilon^2)^2+4\varepsilon^2 a^2}\right]_{t=0}^{t=1} +
\lambda_{\varepsilon,a} {\cal I}_0 (\varepsilon, a)   \\[1ex] &=&
Q_{\varepsilon,a} (1)\sqrt{(1-a^2 +\varepsilon^2)^2+4\varepsilon^2
a^2} - Q_{\varepsilon,a}(0) (\varepsilon^2 +a^2) +
\lambda_{\varepsilon,a} {\cal I}_0 (\varepsilon, a),
\end{eqnarray*}
and the equation for $Q(t)$ and $\lambda$ is
\begin{equation}\label{eqa:2}
(1-t)^k=Q_{\varepsilon,a}^{\prime} (t)\left[ (t-a^2
+\varepsilon^2)^2+4\varepsilon^2 a^2 \right] +Q_{\varepsilon,a}(t)
(t-a^2 +\varepsilon^2) +\lambda_{\varepsilon,a} .
\end{equation}
It is apparent from (\ref{eqa:2}) that $Q(t)$ and $\lambda$ must
be polynomials in $a^2$ and $\varepsilon^2$ and, therefore, in the
limit $\varepsilon\to 0$,
\[
Q_{\varepsilon,a}(t)=Q_{a} (t) + O(\varepsilon^2) \hspace{2ex}
\hbox{and} \hspace{2ex} \lambda_{\varepsilon,a} = \lambda_{a} +
O(\varepsilon^2),
\]
and
\[
(1-t)^k=Q_{a}^{\prime} (t) (t-a^2)^2 +Q_{a}(t) (t-a^2)
+\lambda_{a}.
\]
This equation for $Q_{a}$ and $\lambda_{a}$ can be explicitly
solved, the solution being
\[
\lambda_{a}=(1-a^2)^k \hspace{2ex} \hbox{and} \hspace{2ex}
Q_{a}(t)= \sum_{l=1}^{k} \frac{(-1)^{l}}{l} \begin{pmatrix}
  k \\
  l
\end{pmatrix} (t-a^2)^{l-1} (1-a^2)^{k-l}.
\]
Note that at $Q_{a}(0)$ is a polynomial in $a^2$ of degree $k-1$,
and
\[
Q_{a}(1)= (1-a^2)^{k-1} \sum_{l=1}^{k} \frac{(-1)^{l}}{l}
\begin{pmatrix}
  k \\
  l
\end{pmatrix}= -(1-a^2)^{k-1}\gamma_k,
\]
where $\gamma_k$ is the partial sum of the harmonic series,
\[
\gamma_k = 1+ \frac{1}{2} +\ldots + \frac{1}{k}.
\]

We now turn to ${\cal I}_0 (\varepsilon^2, a^2)$. Recalling the
table integral
\[
\int\frac{dt}{\sqrt{t^2+\alpha^2}}=\ln |t+\sqrt{t^2+\alpha^2}|,
\]
we have
\[
{\cal I}_0 (\varepsilon^2, a^2) =  \ln \frac{1-a^2+\varepsilon^2 +
\sqrt{(1-a^2 +\varepsilon^2)^2+4\varepsilon^2 a^2} }{2
\varepsilon^2}
\]
At $a^2=1$,
\[
{\cal I}_0 (\varepsilon^2, 1) = \ln \frac{\varepsilon^2 +
\sqrt{\varepsilon^4+4\varepsilon^2} }{2 \varepsilon^2}=\ln
\frac{1}{\varepsilon} +O(\varepsilon).
\]
For $a^2\not= 1$,
\[
\sqrt{(1-a^2 +\varepsilon^2)^2+4\varepsilon^2 a^2} = |1-a^2| +
\frac{\varepsilon^2(1+a^2)}{|1-a^2|} +O(\varepsilon^4),
\]
and therefore
\[
{\cal I}_0 (\varepsilon^2, a^2)=
  \begin{cases}
    \displaystyle{\ln \frac{1-a^2}{\varepsilon^2} + O(\varepsilon^2)} & \text{if $a^2<1$},
    \\[1ex]
    \displaystyle{\ln \frac{a^2}{a^2-1} + O(\varepsilon^2)} & \text{if $a^2>1$}.
  \end{cases}
\]
After collecting all relevant terms we arrive at the desired
formula
\[
{\cal I}_k (\varepsilon^2, a^2) = (1-a^2)^k \theta (1-a^2) \ln
\frac{1}{\varepsilon^2}+ \beta (a^2) + q_k (a^2) +O(\varepsilon),
\]
where $\theta$ is the Heaviside step-function,
\[
\theta (x) =
  \begin{cases}
    1 & \text{if $x>0$}, \\
    \frac{1}{2} & \text{if $x=0$},\\
    0 & \text{if $x<0$},
  \end{cases}
\]
$q_k$ is a polynomial of degree $k$ and
\[
\beta (a^2) = \sign (a^2-1) (1-a^2)^k(\gamma_k -\ln |1-a^2|) +
\theta (a^2-1) \ln a^2.
\]
We use the convention according to which the sign function, $\sign
(x)$, vanishes at $x=0$.

\section{Appendix} \label{AppendixB}

In this appendix we derive equation (\ref{eq5:11}). Let
\[
W_n = R
\begin{pmatrix}
  z & \vec{w}^* \\
  \vec{0} & W_{n-1}
\end{pmatrix} R, \hspace{3ex} R= \begin{pmatrix}
  2\vec{q}^*\vec{q}-1& -2\sqrt{1-\vec{q}^*\vec{q}}\vec{q}^* \\
  -2\sqrt{1-\vec{q}^*\vec{q}}\vec{q} & I_{n-1} - 2\vec{q}\vec{q}^*
\end{pmatrix}
\]
When $z, \vec{w}, \vec{\eta}$ and $W_{n-1}$ get infinitesimal
increments $dz, d\vec{w}, d\vec{q}$ and $dW_{n-1}$ the matrix
$W_n$ gets increment
\[
dW_n= dR \begin{pmatrix}
  z & \vec{w}^* \\
  0 & W_{n-1}
\end{pmatrix} R + R \begin{pmatrix}
  dz & d\vec{w}^* \\
  0 & dW_{n-1}
\end{pmatrix}R + R \begin{pmatrix}
  z & \vec{w}^* \\
  0 & W_{n-1}
\end{pmatrix} dR.
\]
Since $R$ is unitary Hermitian, the matrix $RdR$ is
skew-Hermitian, so that
\[
dT=RdR=\begin{pmatrix}
  df & -d\vec{h}^* \\
  d\vec{\sigma} & dT_{n-1}
\end{pmatrix}
\]
for some $f$, $\vec{h}$ and $T_{n-1}$. Also $RdR=-(dR) R$, and it
follows that
\begin{eqnarray}\nonumber
\!\!\!\!\!\!R(dW_n)R\!\!&\!\!=\!\!&\!\!dT\begin{pmatrix}
  z & \vec{w}^* \\
  0 & W_{n-1}
\end{pmatrix} - \begin{pmatrix}
  z & \vec{w}^* \\
  0 & W_{n-1}
\end{pmatrix} dT +  \begin{pmatrix}
  dz & d\vec{w}^* \\ \label{eq5:4}
  0 & dW_{n-1}
\end{pmatrix}\\ \!\!&\!\!=\!\!& \!\!\!\!\begin{pmatrix}
  \vec{w}^*d\vec{h} & \vec{w}^*df \!-\!zd\vec{h}^*\!-\!d\vec{h}^* W_{n-1}\! +\!\vec{w}^*dT_{n-1} \\
  \!(zI\!-\!W_{n-1})d\vec{h} & d\vec{h}
  \vec{w}^*\!+\!dT_{n-1}W_{n-1}\!-\!W_{n-1}dS_{n-1}
\end{pmatrix} \!+\!
 \begin{pmatrix}
  dz & d\vec{w}^* \\
  0 & dW_{n-1}
\end{pmatrix}
\end{eqnarray}
Let $dM=R(dW_n)R$. It is apparent that
\begin{equation}\label{eq5:6}
\prod_{j,k=1}^n d(W_n)_{jk} d\overline{(W_n)_{jk}} =
\prod_{j,k=1}^n dM_{jk} d\overline{M}_{jk}.
\end{equation}
On the other hand, it follows from (\ref{eq5:4}) that
\begin{equation}\label{eq5:7}
\prod_{j,k=1}^n dM_{jk} d\overline{M}_{jk} \!\!=\!\!
|\det(zI-W_{n-1})|^2 \ dz d\overline{z} \ \prod_{j=1}^{n-1} dw_{j}
d\overline{w}_{j}\ \prod_{j=1}^{n-1} dh_{j} d\overline{h}_{j}\
\prod_{j,k=1}^{n-1} d(W_{n-1})_{jk} d\overline{(W_{n-1})}_{jk}.
\end{equation}
To complete our derivation we now compute the Jacobian of the
transformation from $\vec{h}$ to $\vec{q}$. Recall that $d\vec{h}$
is the (2,1)-entry of the matrix $dT=RdR$. A straightforward
computation yields
\begin{equation}\label{eq5:5}
d\vec{h} = (2a+b)(d\vec{q}^*) \vec{q} \vec{q} + a(d\vec{q})
+b\vec{q}\vec{q}^* (d\vec{q}),
\end{equation}
where
\[
a=-2\sqrt{1-\vec{q^*} \vec{q}}, \hspace{2ex} b= \frac{1-
2\vec{q}^* \vec{q}}{\sqrt{1-\vec{q}^* \vec{q}}}.
\]
Equation (\ref{eq5:5}) can be written as
\[
d\vec{h} = (aI+b\vec{q} \vec{q}^*)(d\vec{q}) +
(2a+b)\vec{q}\vec{q}^T (d\overline{\vec{q}}),
\]
and, therefore,
\[
\begin{pmatrix}
  d\vec{q} \\
  d\overline{\vec{q}}
\end{pmatrix} =
\begin{pmatrix}
   aI + b\vec{q} \vec{q}^*  &  (2a+b)\vec{q} \vec{q}^T \\
    (2a+b)\overline{\vec{q}} \vec{q}^* &  aI+ b \overline{\vec{q}} \vec{q}^T \
   \end{pmatrix}
\begin{pmatrix}
  d\vec{q} \\
  d\overline{\vec{q}}
\end{pmatrix}.
\]
It now follows that
\begin{equation}\label{eq5:8}
\prod_{j=1}^{n-1} d h_j d\overline{h}_j = \det (aI+L)
\prod_{j=1}^{n-1} d q_j d\overline{q}_j
\end{equation}
where $L$ is the $2(n-1)\times 2(n-1)$ matrix
\[
\begin{pmatrix}
   b\vec{q} \vec{q}^*  &  (2a+b)\vec{q} \vec{q}^T \\
    (2a+b)\overline{\vec{q}} \vec{q}^* &  b \overline{\vec{q}} \vec{q}^T
   \end{pmatrix}.
\]
If we find the eigenvalues of $L$, we shall know $\det (aI + L)$.

To solve the eigenvalue problem for $L$, we observe that if
$(\vec{f}, \vec{g})^T$ is an eigenvector of $L$ then
\[
\left\{\begin{array}{l c l}b(\vec{q}^* \vec{f}) \vec{q}  + (2a+b)
(\vec{q}^T\vec{g}) \vec{q}  &= &\lambda \vec{f}\\ (2a+b)
(\vec{q}^* \vec{f}) \overline{\vec{q}}  + b (\vec{q}^T \vec{g})
\overline{\vec{q}}  &=& \lambda \vec{g}
\end{array} \right.
\]
for some $\lambda$. If $\lambda \not=0$ we must have $\vec{f}=c_1
\vec{q}$ and $\vec{g}=c_2 \overline{\vec{q}}$ for some $c_1$ and
$c_2$, and
\[
\left\{\begin{array}{l c l}b(\vec{q}^* \vec{q}) c_1 + (2a+b)
(\vec{q}^*\vec{q})c_2  &= &\lambda c_1\\ (2a+b) (\vec{q}^*
\vec{q})c_1   + b (\vec{q}^* \vec{q})c_2 &=& \lambda c_2
\end{array} \right.
\]
This reduced eigenvalue problem yields the two non-zero
eigenvalues of $L$,
\[
\lambda_1 = -2a \vec{q}^*\vec{q}\hspace{1ex} \hbox{and}
\hspace{1ex} \lambda_2= 2(a+b)\vec{q}^*\vec{q}.
\]
It is now apparent that $\lambda=0$ is an eigenvalue of $L$ of
multiplicity $2(n-2)$. This fact can be verified independently of
the eigenvalue count by observing that for any vector $\vec{u}$
which is orthogonal to $\vec{q}$,
\[
L\begin{pmatrix}
   \vec{u} \\
    0 \
   \end{pmatrix} =0 \hspace{1ex} \hbox{and} \hspace{1ex} L\begin{pmatrix}
   0 \\
   \overline{\vec{u}} \
   \end{pmatrix} =0.
\]
It follows now that
\begin{equation}\label{eq5:9}
\det (aI+ L)= a^{2(n-2)}(a+\lambda_1)(a+\lambda_2) =
(-2)^{2n-2}(1-\vec{q}^*\vec{q})^{n-2} (1-2\vec{q}^*\vec{q}).
\end{equation}
Collecting (\ref{eq5:6}) - (\ref{eq5:7}) and (\ref{eq5:8}) -
(\ref{eq5:9}), one arrives at $(\ref{eq5:11})$.


\begin{thebibliography}{99}
\bibitem{AV} Akemann, G., Vernizzi G.: Characteristic Polynomials of
Complex Random Matrix Models. Nucl.Phys. B {\bf 660}, 532--556
(2003).

\bibitem{AP} Akemann, G., Pottier, A. Ratios of characteristic
polynomials in complex matrix models.: J.Phys. A: Math and General
{\bf 37}, L453--L460 (2004).

\bibitem{AS} Andreev, A.V., Simons, B.D.: Correlators of
spectral determinants in quantum chaos. Phys. Rev. Lett. {\bf 75},
2304--2307 (1995).

\bibitem{Balantekin} Balantekin, A.B.: Character expansions,
Itzykson-Zuber integrals, and the QCD partition function.  Phys.
Rev. D(3) {\bf 62}, 085017--085023 (2000).

\bibitem{BDS} Baik, J.,  Deift, P. Strahov, E.: Products and ratios of
characteristic polynomials of random Hermitian matrices. J. Math.
Phys. {\bf 44}, 3657--3670  (2003) .

\bibitem{Berezin2} Berezin, F.A.: Some remarks on the Wigner
distribution (in Russian). Teor. Mat. Fiz. {\bf 17}, 305--318
(1973).

\bibitem{Berezin1} Berezin, F.A.: Quantization in complex
symmetric spaces. Math. USSR - Izv. {\bf 9}, 341--379 (1976).

\bibitem{BL} Biane, Ph., Lehner, F.: Computation of some examples
of Brown's spectral measure in free probability. ESI Preprint No.
823, 27 pages.

\bibitem{BOS}
Borodin, A., Olshanski, G., Strahov, E.: Giambelli compatible
point processes. E-preprint ArXiv:math-ph/0505021.

\bibitem{BS}
Borodin, A., Strahov, E.: Averages of characteristic polynomials
in Random Matrix Theory.  Commun. Pure and Applied Math., {\bf 59}
(2), 161--253 (2006).

\bibitem{BH} Brezin, E., Hikami, S.: Characteristic polynomials
of random matrices. Commun. Math. Phys. {\bf 214}, 111--135
(2000).

\bibitem{BG} Bump, D., Gamburd, A.: On the average of
characteristic polynomials from classical groups. Comm. Math.
Phys., in press.

\bibitem{CFKRS}
Conrey, J.B., Farmer, D.W., Keating, J.P., Rubinstein, M.O.,
Snaith, N.C.:  Autocorrelation of random matrix polynomials.
Commun. Math. Phys. 237, 365--395 (2003).

\bibitem{CFS}
Conrey, J.B., Forrester, P.J., Snaith, N.C.:  Averages of ratios
of characteristic polynomials for the compact classical groups,
IMRN {\bf 7}, 397--431 (2005).

\bibitem{CFZ} Conrey, J.B., Farmer, D.W., Zirnbauer, M.R.:
Howe pairs, supersymmetry, and ratios of random characteristic
polynomials for the unitary groups U(N). E-preprint
ArXiv:math-ph/0511024.

\bibitem{DG} Diaconis, P., Gamburd, A.: Random matrices, magic squares
and matching polynomials.:  Electron. J. Combin.  {\bf 11}, no. 2
(2004/05), research paper 2, 26 pp.

\bibitem{E} Edelman, A.: The probability that a random real
gaussian matrix has $k$ real eigenvalues, related distributions,
and the Cirular law. J. Multiv. Anal. {\bf 60}, 203--232 (1997).

\bibitem{EKS} Edelman, A., Kostlan, E., Shub, M.: How many
eigenvalues of a random matrix are real? J. Amer. Math. Soc. {\bf
7}, 247--267  (1994).

\bibitem{FZ} Feinberg, J. and Zee, A.,
Non-Gaussian Non-Hermitean Random Matrix Theory: phase transitions
and addition formalism. Nucl.Phys. {\bf B501}, 643--669 (1997).

\bibitem{FSZ} Feinberg, J., Scalettar, R., Zee, A.:
"Single Ring Theorem" and the Disk-Annulus Phase Transition.
J.Math.Phys. {\bf 42}, 5718--5740 (2001).


\bibitem{Fyo} Fyodorov, Y.V.: Negative moments of characteristic polynomials of
random matrices: Ingham-Siegel integral as an alternative to
Hubbard-Stratonovich transformation. Nuclear Phys. B {\bf 621},
643--674 (2002).

\bibitem{FA} Fyodorov, Y.V.,  Akemann, G.:
On the supersymmetric partition function in QCD-inspired random
matrix models, JETP Lett. {\bf 77}, 438--441 (2003).


\bibitem{FKh}  Fyodorov, Y.V., Khoruzhenko, B.A.: Systematic analytical
approach to correlation functions of resonances in quantum chaotic
scattering. Phys. Rev. Let. {\bf 83}, 65--68  (1999).

\bibitem{FSom1} Fyodorov, Y.V., Sommers, H.-J.: Statistics of
resonance poles, phase shifts and time delays in quantum chaotic
scattering: Random matrix approach for systems with broken
time-reversal invariance. J. Math. Phys. {\bf 38}, 1918--1981
(1997).

\bibitem{FSom2} Fyodorov, Y.V., Sommers, H.-J.: Random matrices
close to Hermitian or unitary: overview of methods and results. J.
Phys. A {\bf 36}, 3303--3347 (2003).

\bibitem{FS1} Fyodorov, Y.V., Strahov E.: An exact formula for
general spectral correlation function of random Hermitian
matrices. J. Phys. A  {\bf 36}, 3203--3213 (2003).

\bibitem{FStra}
Fyodorov, Y.V., Strahov, E.: Characteristic polynomials of random
Hermitian matrices and Duistermaat-Heckman localisation on
non-compact K\"{a}hler Manifolds. Nucl.Phys. {\bf B630}, 453--491
(2002).


\bibitem{HL}  Haagerup, U., Larsen, F.: Brown's spectral
distribution measure for $R$-diagonal elements in finite von
Neumann algebras. J. Fun. Analysis {\bf 176}, 331--367 (2000).

\bibitem{HJV} Halasz, M.A., Jackson, A.D.,  Verbaarschot, J.J.M.:
Fermion determinants in matrix models of QCD at nonzero chemical
potential. Phys. Rev. {\bf D56},  5140--5152 (1997).

\bibitem{Hua} Hua, L.K.: Harmonic Analysis of Functions of Several
Complex variables in the Classical Domains. Amer. Math. Soc:
Providence, Rhode Island, 1963.

\bibitem{G} Ginibre, J.: Statistical Ensembles of Complex,
Quaternion, and Real Matrices. J. Math. Phys. {\bf 6}, 440--449
(1964).

\bibitem{GR} Gradshtein, I.S., Ryzhik, I.M.: Table of
Integrals, Series, and Products, 5th ed., A. Jeffrey, editor:
Academic Press, 1994.

\bibitem{Kad} Kadell, K.W.J.: The Selberg-Jack symmetric
functions. Advances in Math. {\bf 130}, 33--102 (1997).

\bibitem{Ka} Kaneko, J.: Selberg integrals and hypergeometric
functions associated with Jack polynomials. SIAM Jour. Math. Anal.
{\bf 24}, 1086--1110 (1993).

\bibitem{KS1} Keating, J.P.,  Snaith, N.C.: Random matrix
theory and $\zeta (1/2 + it))$. Comm. Math. Phys. {\bf 214},
57--89 (2000).

\bibitem{KS2} Keating, J. P., Snaith, N. C.: Random matrix
theory and $L$-functions at $s=1/2$. Comm. Math. Phys. {\bf 214},
91--110 (2000).

\bibitem{Macdonald} Macdonald, I.G., Symmetric Functions and Hall
Polynomials. 2nd ed., Clarendon Press: Oxford, 1995.

\bibitem{Mehta} Mehta, M.L.: Random matrices. 3rd ed, Elsevier/Academic Press:
Amsterdam, 2004.

\bibitem{O1} Orlov, A. Yu.: New Solvable Matrix
Integrals. In Proceedings of 6th International Workshop on
Conformal Field Theory and Integrable Models. Internat. J. Modern
Phys. A {\bf 19},  May, suppl., 276--293 (2004).

\bibitem{PS} Polya, G., Szego, G.: Problems and Theorems in Analysis.
Vol.2, Chapter 6, Problem 67. Springer: New-York 1978.

\bibitem{SW} Schlittgen, B., Wettig T.: Generalizations of some integrals
over the unitary group. J. Phys. A {\bf 36}, 3195--3202 (2003).

\bibitem{SV} Shuryak, E.V., Verbaarschot, J.J.M.: Random matrix theory and spectral
sum rules for the Dirac operator in QCD.  Nucl. Phys. {\bf A560},
306--320 (1993).

\bibitem{Strahov} Strahov, E.: Moments of characteristic polynomials enumerate
two-rowed lexicographic arrays. Electronic Journal of
Combinatorics, {\bf 10}, R24 (2003).


\bibitem{Trotter} Trotter, H.F.: Eigenvalue distributions of large
Hermitian matrices: Wigner semicircle and a theorem of Kac,
Murdock, and Szego. Adv. Math. {\bf 54}, 67--82 (1984).

\bibitem{V1} Verbaarschot, J. J. M.: Spectrum of the QCD Dirac
Operator and Chiral Random Matrix Theory. Phys. Rev. Lett. {\bf 72
}, 2531--2533 (1994).

\bibitem{V2} Verbaarschot, J. J. M.: QCD, chiral random matrix
theory and integrability. E-preprint arXiv:hep-th/0502029.

\bibitem{Wilkinson} Wilkinson, J. H.: The Algebraic Eigenvalue
Problem, Clarendon Press: Oxford, 1965.

\bibitem{Z} Zirnbauer, M.R.: Supersymmetry for systems with unitary disorder: circular ensembles.
J. Phys. A {\bf 29}, 7113--7136  (1996).

\bibitem{ZS} Zyczkowski, K., Sommers, H.-J.: Truncations of random unitary matrices.
J. Phys. A {\bf 33}, 2045--2058  (2000).
\end{thebibliography}
\end{document}